\documentclass{aas}
\usepackage{graphicx}
\usepackage{tabularx}
\usepackage{longtable}
\usepackage{rotating}

\received{  }
\revised{}
\accepted{}
\submitjournal{ }

\shorttitle{Search for hot subdwarf stars in Gaia DR2 }
\shortauthors{Lei et al.}

\begin{document}

\title{New hot subdwarf stars identified in Gaia DR2 with LAMOST DR5 spectra - II.  }

\correspondingauthor{Gang Zhao}
\email{gzhao@nao.cas.cn}

\correspondingauthor{Zhenxin Lei}
\email{zxlei@nao.cas.cn}

\author{Zhenxin Lei }
\affiliation{Key Laboratory of Optical Astronomy, National Astronomical Observatories, Chinese Academy of Sciences, Beijing 100012, China\\
}
\affiliation{College of Science, Shaoyang University, Shaoyang 422000, China\\}

\author{Jingkun Zhao }
\affiliation{Key Laboratory of Optical Astronomy, National Astronomical Observatories, Chinese Academy of Sciences, Beijing 100012, China\\
}

\author{P\'eter N\'emeth}
\affiliation{ Astronomical Institute of the Czech Academy of Sciences, CZ-251\,65, Ond\v{r}ejov, Czech Republic\\}
\affiliation{ Astroserver.org, 8533 Malomsok, Hungary\\
}

\author{Gang Zhao }
\affiliation{Key Laboratory of Optical Astronomy, National Astronomical Observatories, Chinese Academy of Sciences, Beijing 100012, China\\
}



\begin{abstract}
388 hot subdwarf stars have been identified by using the Hertzsprung-Russell (HR) diagram 
built from the second data release (DR2) of the Gaia mission. By analyzing their 
observed LAMOST spectra,  we characterized 186 sdB, 
73 He-sdOB, 65 sdOB,  45 sdO, 12 He-sdO and 7 He-sdB stars.  
The atmospheric parameters of these stars (e.g., $T_\mathrm{eff}$, $\mathrm{log}\ g$, 
$\mathrm{log}(n\mathrm{He}/n\mathrm{H})$) are obtained by fitting 
the hydrogen (H) and helium (He) line profiles with synthetic spectra 
calculated from non-Local Thermodynamic Equilibrium (non-LTE) model atmospheres. 
Among these stars, we have 135 new identified hot subdwarfs which have not been 
cataloged before. Although 253 stars appear in the catalog by Geier et al. (2017) , but only 
91 of them have atmospheric parameters. 
Together with the 294 hot subdwarf stars found by Lei et al. (2018), we identified 682 
hot subdwarf stars in total by using the Gaia HR-diagram and  LAMOST spectra. 
These results demonstrate the efficiency of our method 
to combine large surveys to search for hot subdwarf stars. 
We found a distinct gap in our He-sdOB stars based on their He abundance, which is also 
presented in extreme horizontal branch (EHB) stars of the globular cluster (GC) $\omega$ Cen. The  
number fraction of the sample size  
for the two sub-groups is very different between the two counterparts. However, the distinct gap between 
the H-sdB stars and He-sdOB stars in $\omega$ Cen is not visible in our sample.  
More interestingly, the He-sdB population with the highest He abundance in our sample is completely 
missing in $\omega$ Cen. The discrepancy  between our field hot subdwarf stars and 
the EHB stas in $\omega$ Cen indicate different origins for the two counterparts. 

\end{abstract}



\section{Introduction} \label{sec:intro}
Hot subdwarf (e.g., spectral type sdB, sdO and other sub-types of) stars 
are the exposed helium burning cores  of  red giant branch (RGB) stars. They have low stellar masses 
around 0.5 M$_{\odot}$ and very thin H-rich envelopes (e.g., $\le$ 0.01 M$_{\odot}$;  
Heber 2009). Hot subdwarf stars play very important roles in many aspects of 
astrophysics. Studies on the formation of these special blue stars 
will vastly improve our understanding on  stellar structure and evolution 
of low mass stars (Han et al. 2002, 2003). Pulsating hot subdwarfs 
studied by  asteroseismology give insights into  their interior  
structures (Kawaler et al. 2010; Charpinet et al. 2011; Baran et al. 2012; 
Battich et al. 2018;  Zong et al. 2018). 
Furthermore, the variety  of surface chemical compositions in hot subdwarfs stars 
make them good samples to study atmotic diffusion processes 
(Hu et al. 2009, 2010, 2011; Naslim et al. 2013; Moehler et al. 2014; Jeffery et al. 2017;  
N\'emeth 2017;  Byrne et al. 2018).  
Study on the counterparts of field hot subdwarfs 
in GCs (e.g., EHB stars) can help us 
understand the formation history and evolution processes of these 
old populations (Lei et al. 2015, 2016; Latour et al. 2014, 2018). 
For a recent review on hot subdwarf stars see Heber (2009, 2016). 

Binary evolution is considered as the main formation channel 
of hot subdwarf stars, since about half of these stars are found 
in close binaries (Maxted et al. 2001; Napiwotzki et al. 2004; Copperwheat et al. 2011). 
On the theoretical side, Han et al. (2002, 2003) conducted detailed binary evolution and found 
that Roche lobe overflow (RLOF), common 
envelope (CE) ejection and merger of two He white dwarfs (WDs) can form sdB stars 
in binary systems with their properties matching well with the observations. 
Recently, Kawka et al. (2015) have reviewed the properties of hot 
subdwarf binary population and measured the orbital parameters of seven close binaries comprising 
hot subdwarf stars selected from the Galaxy Evolution Explorer ({\it GALEX},  
Morrissey et al. 2007) survey, while in a similar study Kupfer et al. (2015) presented the 
orbital and atmospheric parameters and put constraints on the nature of the 
companions of 12 close sdB stars found in 
the project Massive Unseen Companions to Hot Faint Underluminous Stars from SDSS (MUCHFUSS). 
Zhang \& Jeffery (2012) found that the merger of two He-WDs can reproduce the 
observed distribution of He-rich hot subdwarf stars in terms of effective 
temperatures, surface gravities, nitrogen and carbon abundances. Moreover, 
Zhang et al. (2017) found that the mergers of He-WDs with low mass 
main-sequence (MS) stars can form intermediate He-rich hot subdwarf stars. 
Recent progress in this field can be found in Chen et al. (2013), Xiong et al. 2017, \
Wu et al. 2018 and Vos et al. 2019. 

Hot subdwarf stars  consist of several sub-types according to 
their spectral features, such as sdB, sdO, sdOB, He-sdB, He-sdO and He-sdOB 
(Moehler et al. 1990; Geier et al. 2017a). Moehler et al. (1990)  
introduced a detailed spectral classification scheme for hot subdwarf stars, which is also  
widely used today. In this scheme, the spectral features of hot subdwarfs 
with different sub-types were described with details, including their 
differences to B type main-sequence stars, blue horizontal branch (BHB) stars 
and white dwarfs (WDs).  Recently, Drilling et al. (2013) defined a more sophisticated MK-like classification scheme 
for hot subdwarf stars. They found that a spectral class, 
a luminosity class and a helium class is necessary to classify 
these hot stars. Moreover, they also give a preliminary calibration 
between the new spectral classification and atmospheric parameters 
(eg., effective temperature, surface gravity, and surface helium-to-hydrogen abundance ratio).  

Spectral analysis is the ordinary method to 
obtain atmospheric parameters (e.g., effective temperature, 
gravity) and surface chemical abundances of hot subdwarf stars. 
Lisker et al. (2005) analyzed sdB stars found in 
the ESO Supernova Ia Progenitor Survey (ESO-SPY, Napiwotzki et al. 2001) 
by using metal line-blanketed LTE models of solar composition
and the LINFOR program for spectrum synthesis (Heber et al. 
1999, 2000). On the other hand, 
Stroeer et al. (2007) analyzed O type subluminous stars of the same sample 
by using non-LTE models with H/He composition. These models 
were calculated with the TMAP package (Werner \& Dreizler 1999; Rauch \&
Deetjen 2003; Werner et al. 2003). Vennes et al. (2011) 
computed a grid of non-LTE model atmospheres
and synthetic spectra by using the codes {\sc Tlusty-Synspec} 
(Hubeny \& Lanz 1995; Lanz \& Hubeny 1995) to 
analyze hot subdwarf stars in the {\it GALEX}   
survey (Morrissey et al. 2007). Moreover, N\'emeth et al. (2012) 
obtained the atmospheric parameters of 166 sdB/O stars found in the \textit {GALEX} survey 
by using H/He/CNO non-LTE model atmospheres with H/He/CNO composition calculated with 
{\sc Tlusty-Synspec}. For recent progress in this field 
see Moni Bidin et al. (2012); Latour et al. (2013, 2015, 2016), etc. 

Due to the significance of hot subdwarfs in astrophysics, many new  
hot subdwarf stars were discovered in large spectral surveys, 
such as Kepler ($\varnothing$stensen et al. 2010),  
{\it GALEX} (Vennes et al. 2011; N\'emeth et al. 2012; Kawka et al. 2015), 
the Sloan Digital Sky Survey (SDSS, Geier et al. 2015; Kepler et al. 2015, 2016) 
and the Large Sky Area Multi-Object Fibre Spectroscopic Telescope (LAMOST) survey 
(Luo et al. 2016). Geier et al. (2017a) compiled a large hot subdwarf catalog   
which contains 5613 objects by retrieving known hot subdwarfs and candidates 
from the literature and unpublished databases. In this catalog, many useful 
information of the stars are listed, such as multi-band photometry, 
proper motions, classifications, atmospheric parameters, etc.     
Furthermore, Geier et al. (2019) compiled an all-sky catalog of 39\,800 hot 
subluminous star candidates, which were selected from the Gaia DR2 database (Gaia Collaboration et al. 2018a) 
by means of colour, absolute magnitude and reduced proper motion cuts. 
The majority of the candidates in this catalog  are expected to be 
hot subdwarf stars, and it can be used as an input catalog for 
future  photometric and spectroscopic surveys. 

With the DR2 of the Gaia mission (Gaia Collaboration et al. 2018a), 
we expect a large number of new hot subdwarf stars  to be uncovered. 
Gaia DR2 provide accurate astrometry (e.g., parallaxes, proper motions) 
and photometry for about 1.3 billion sources over the full sky. With these 
information, an HR-diagram for a huge numbers of stars can be built easily, which 
provides  a very convenient tool to study  stars at different evolutionary stages, 
including hot subdwarf stars. LAMOST is a Chinese national scientific research facility operated 
by the National Astronomical Observatories,  Chinese Academy of Sciences. It has a specially designed  reflecting Schmidt telescope with 4000 fibers in a field of view of 20 deg$^{2}$ in the sky 
(Cui et al. 2012; Zhao et al. 2006, 2012). LAMOST finished its 
pilot survey in 2012 and the first-five-years regular survey in 2017, respectively. The data 
from both  the two surveys make up the fifth data release (DR5) of the LAMOST, in which 
spectra in the optical band (e.g., 3690-9100$\mathrm{\AA}$) for 
8\,171\,443 stars, 153\,090 galaxies, 51\,133 quasars and 642\,178 unknown objects have been 
obtained. This large spectral survey also provide us huge opportunities to analyze the spectra 
of many interesting objects, such as hot subdwarf stars. 

Lei et al. (2018, hereafter Paper I) selected 734 
hot subdwarf candidates from the HR-diagram built by Gaia Collaboration et al. (2018b). 
After analyzing the corresponding  LAMOST spectra, they identified 294 
hot subdwarf stars in their sample, which demonstrated an efficient and powerful 
method to search for hot subdwarf stars by combining the Gaia database and the 
LAMOST spectral database. However, to see the different structures  
clearly, the HR-diagram used to select hot subdwarf 
candidates in Paper I was built by the objects which are strictly filtered out from 
the database of Gaia DR2 (for the detailed filters see Gaia Collaboration et al. 2018b and Paper I ), 
without considering the completeness of the sample. 
Actually, only 65\,921\,112 (e.g., about 33\%) stars 
were selected from the whole 1.7 billion Gaia DR2 sources to build the HR-diagram
by Gaia Collaboration et al. (2018b) , and only 734 of the selected candidates  
have LAMOST spectra, in which 490 spectra have good quality for further spectral analysis. 
It means that many hot subdwarf candidates were not included in the HR-diagram used in Paper I  
due to the strict data selection. Therefore, to recover the maximum number of 
hot subdwarfs by combining the Gaia DR2 database with the LAMOST DR5 database,  
we built a new HR-diagram by using  all the objects from cross-matching the whole Gaia DR2 database and the whole 
LAMOST DR5 database, without any  filters.  
Then we selected the hot subdwarf candidates in the new HR-diagram. This 
method can conserve most of the hot subdwarf stars which were not included in Paper I. 
This paper is structured as follows. In Section 2, we described our new candidates selection process. Spectral 
analysis and classification is presented in Section 3. Our results are shown in Section 4. 
Finally, a discussion and a summary  are given in Section 5 and 6, respectively.

\section{Target selection}
\begin{figure}
\centering

\begin{minipage}[c]{0.40\textwidth}
\includegraphics [width=70mm]{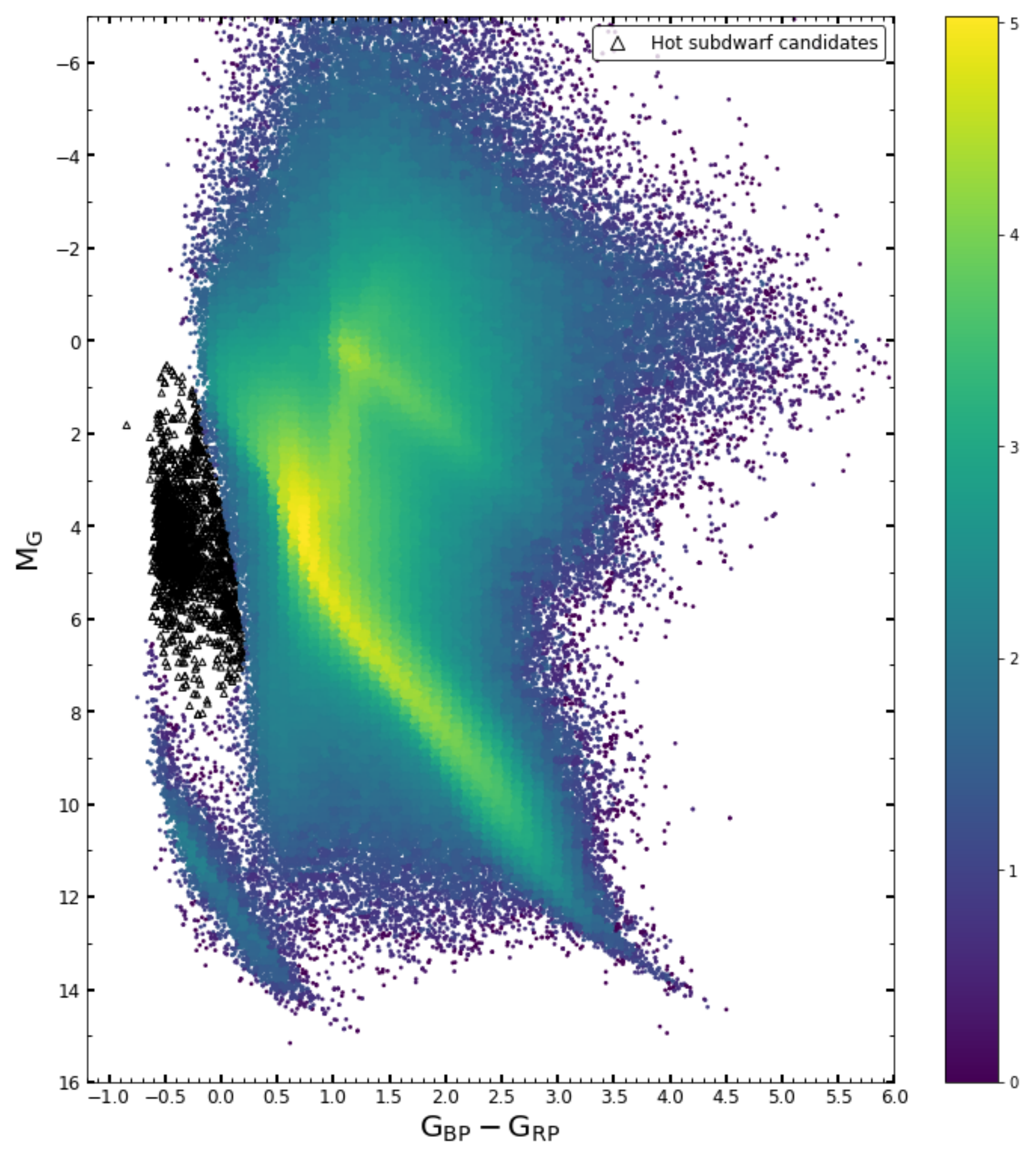}
\centerline{(a) }
\end{minipage}%
\begin{minipage}[c]{0.40\textwidth}
\includegraphics [width=70mm]{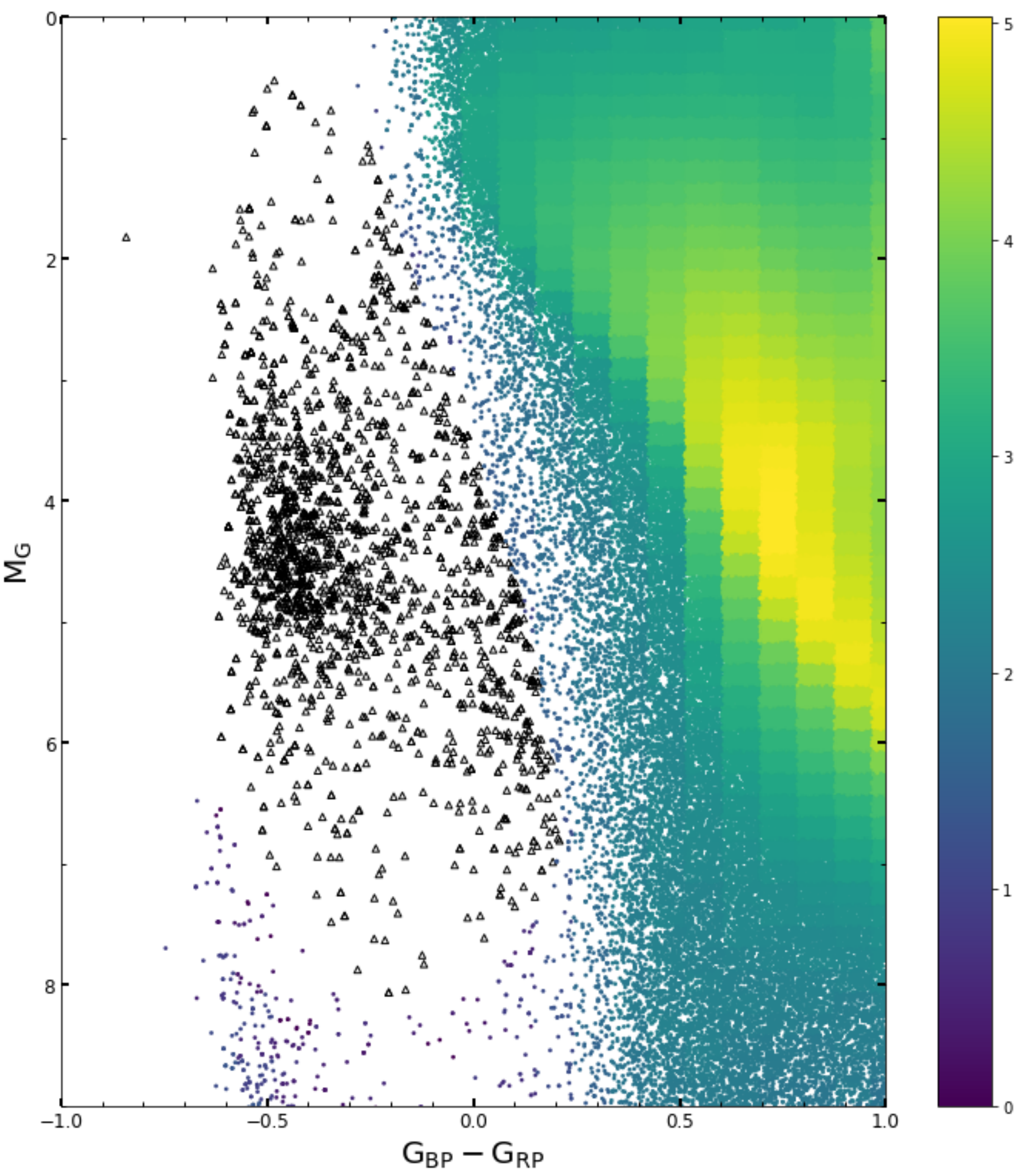} 
\centerline{(b) }
\end{minipage}%

\caption{Panel (a): HR-diagram built with the objects 
filtered out by cross-matching Gaia DR2 database with LAMOST 
DR5 database (totally 8\,852\,848 objects), black triangles are the 2074 hot subdwarf candidates 
selected visually. The color bar represents the  number of objects in 
each bin on a logarithmic scale.  Panel (b): Magnified drawing for  
the region of hot subdwarf candidates. }
\end{figure}

We cross-matched the whole Gaia DR2 data with the LAMOST DR5 catalog, and found 
 8\,852\,848 common objects.  The HR-diagram was built by using the Gaia G$_{BP}$--G$_{RP}$ color and 
 absolute magnitude of the Gaia $G$ band (i.e., $M_{G}$) for all the common objects, 
 where their parallaxes are available and not negative. $M_{G}$ is calculated by the 
 following equation:
 \begin{equation}
 M_{G}=G-5\mathrm{log}_{10}(1000/\varpi)+5, 
 \end{equation}
where $\varpi$ is the parallax in milliarcseconds (mas) and $G$ is the apparent magnitude in the Gaia $G$ band. 
To include the maximum number of hot subdwarfs in our sample, we did not apply any filters  when building 
the HR-diagram. Since the candidates selected from this step will be fitted by synthetic spectra, 
no extinctions are considered in our HR-diagram. 
These measures ensure that the new selection method obtains a larger sample of candidates than the one in Paper I. 

Panel (a) in Fig 1 shows the new HR-diagram. Due to the influences of extinctions, the main-sequence 
(MS) appears much wider, and the red giant branch (RGB)  can not be distinguished from the MS. Fortunately, 
the white dwarf (WD) sequence and hot subdwarf sequence separate more clearly  
from the MS in the HR-diagram due to their much bluer colors than the majority of MS stars. 
The black triangles  in Panel (a) are the candidates we selected visually 
around the hot subdwarf regions. To include hot subdwarf stars as many as possible in our 
sample, we extended our selection region very close to the  left of the wide MS. Because some real 
hot subdwarf stars could settle into these regions due to large extinctions, or some hot subdwarf binaries 
with low mass companions also would locate in these areas. 
The magnified area for the candidates selection is showed in Panel (b) of Fig 1 for clarity. 

Using the method described above, we totally selected 2\,074 candidates in the HR-diagram. 
As we expected, this sample is about 3 times bigger  than the one selected in Paper I \
(e.g., 734 candidates with LAMOST spectra in Paper I). 
After removing the objects we had analyzed in Paper I, we have 1431 objects left in our sample. 
We also removed the objects with the signal to noise ratio (SNR) in $u$ band less than 10 to 
guarantee a good quality of spectral analysis in our follow up study, which reduced the sample to 592 objects.  
After removing double-lined spectroscopic binaries, and spectra with 
obvious spectral contamination from nearby cool stars  
(e.g., Mg I triplet lines at 5170 $\mathrm{\AA}$ and/or Ca II triplet lines at 8650 $\mathrm{\AA}$), 
we finally selected 441 candidates suitable for a  spectral analysis. 

\begin{figure}
\centering
\includegraphics [width=120mm]{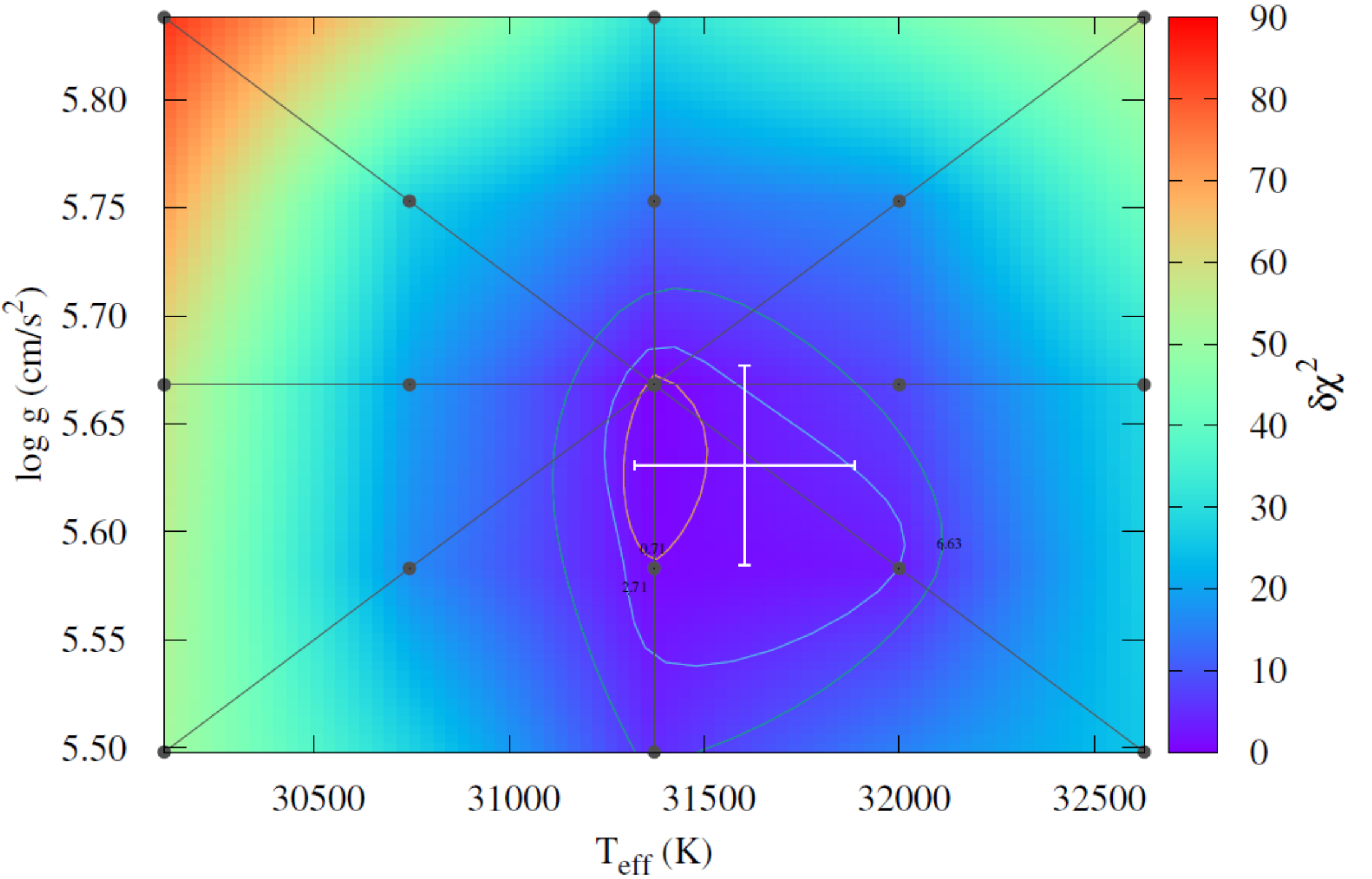}
\caption{$T_{\rm eff}$-$\log{g}$ correlation and two dimensional error determination
for a sdB star (LAMOST obsid: 223907015). The color bar shows the chi-square 
variations with the parameters. The contours show the confidence intervals for 60, 90
and 99\%. See the text for details.} 
\end{figure}

\section{Spectral analysis and classification} 

\begin{figure}
\centering

\begin{minipage}[c]{0.42\textwidth}
\includegraphics [width=75mm]{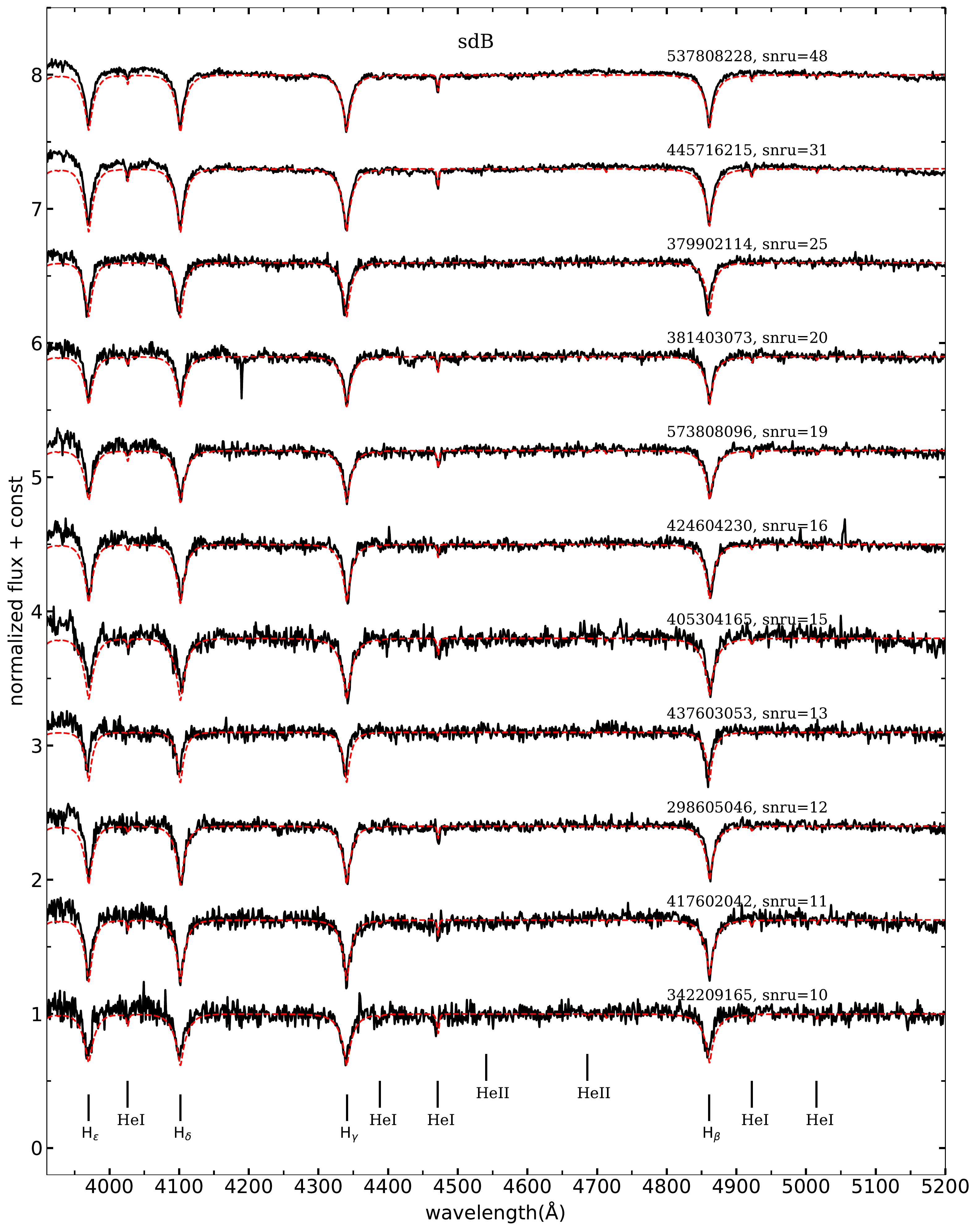}
\centerline{(a) }
\end{minipage}%
\begin{minipage}[c]{0.42\textwidth}
\includegraphics [width=75mm]{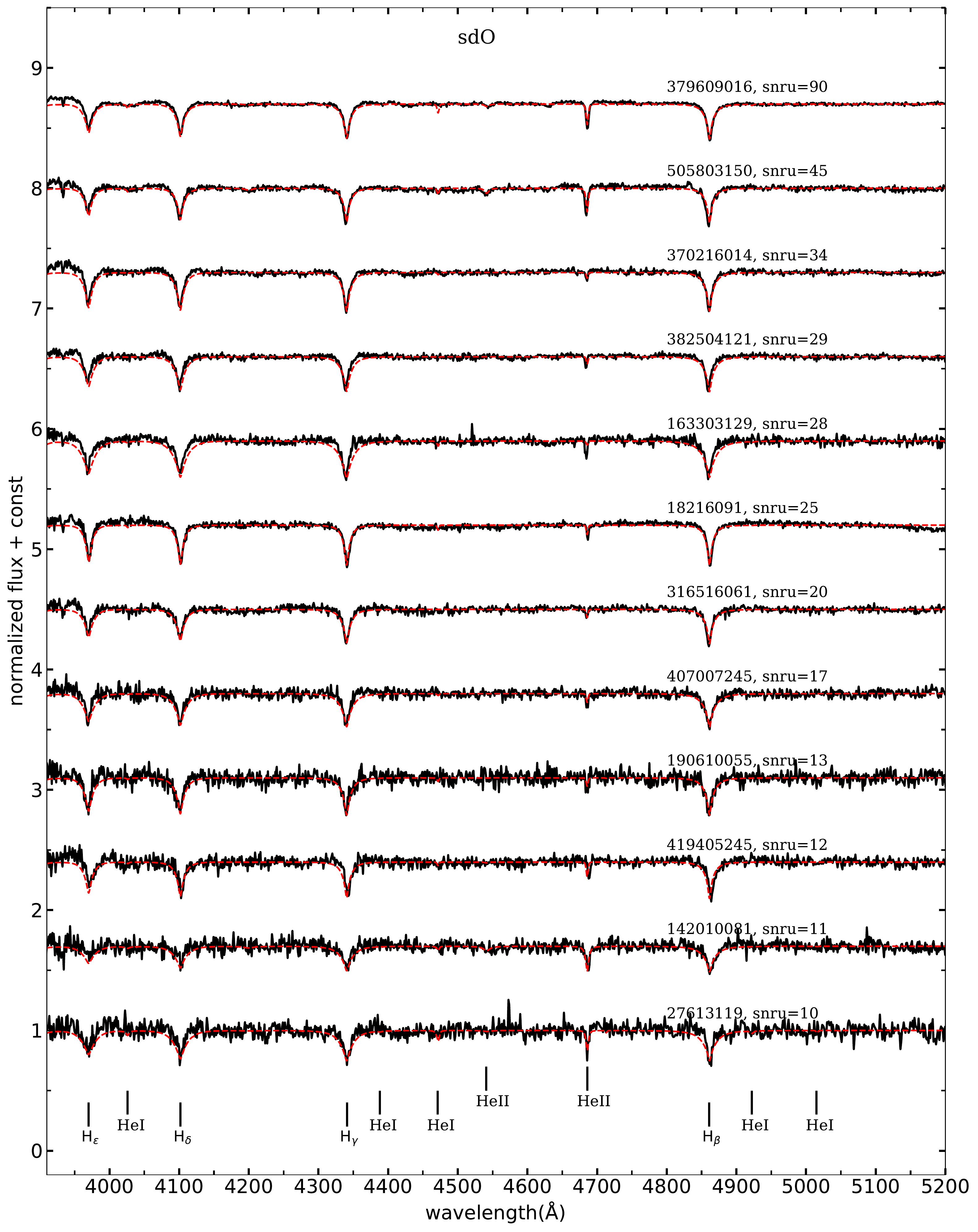} 
\centerline{(b) }
\end{minipage}%
\begin{minipage}[c]{0.42\textwidth}
\includegraphics [width=75mm]{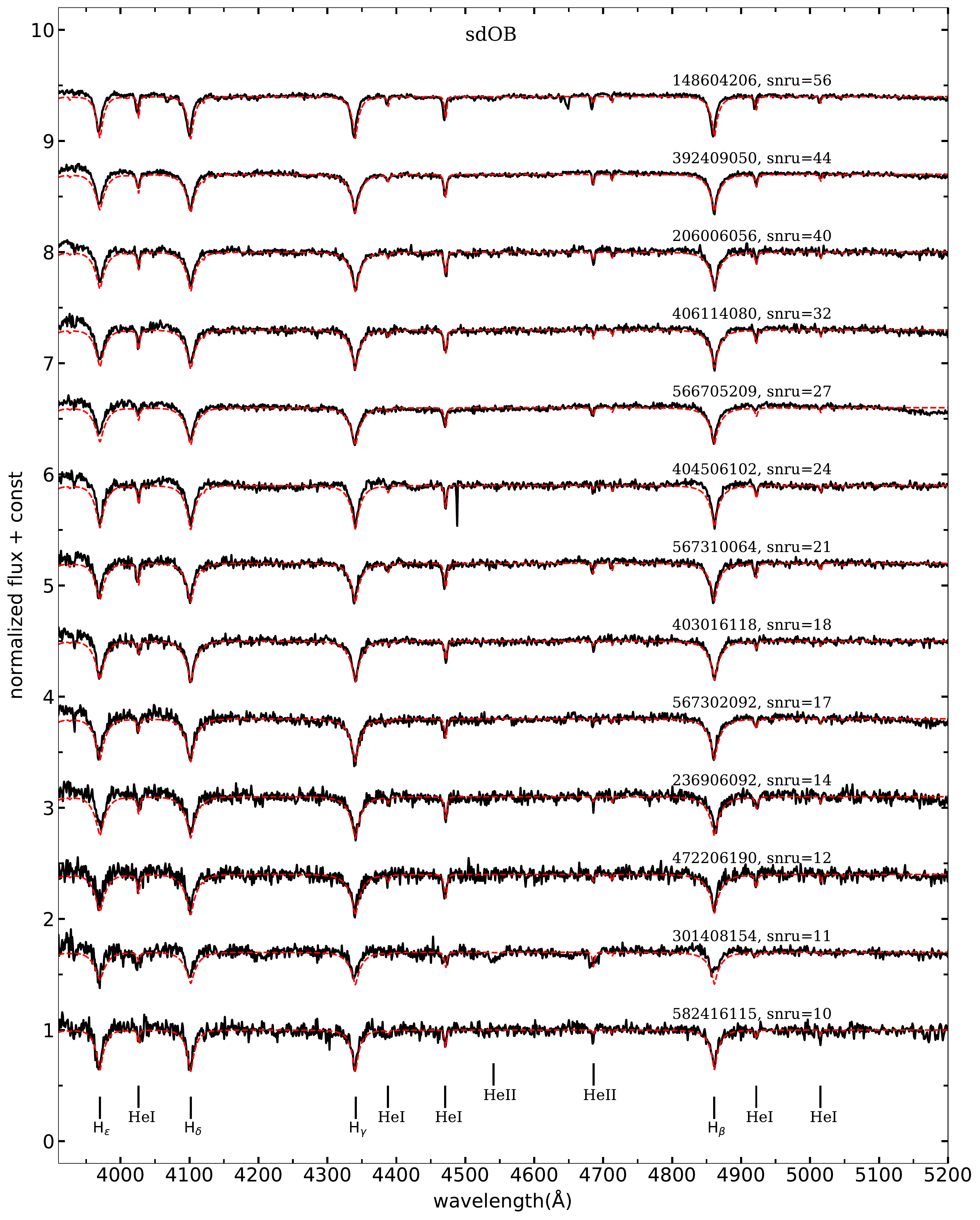} 
\centerline{(c) }
\end{minipage}%
\begin{minipage}[c]{0.42\textwidth}
\includegraphics [width=75mm]{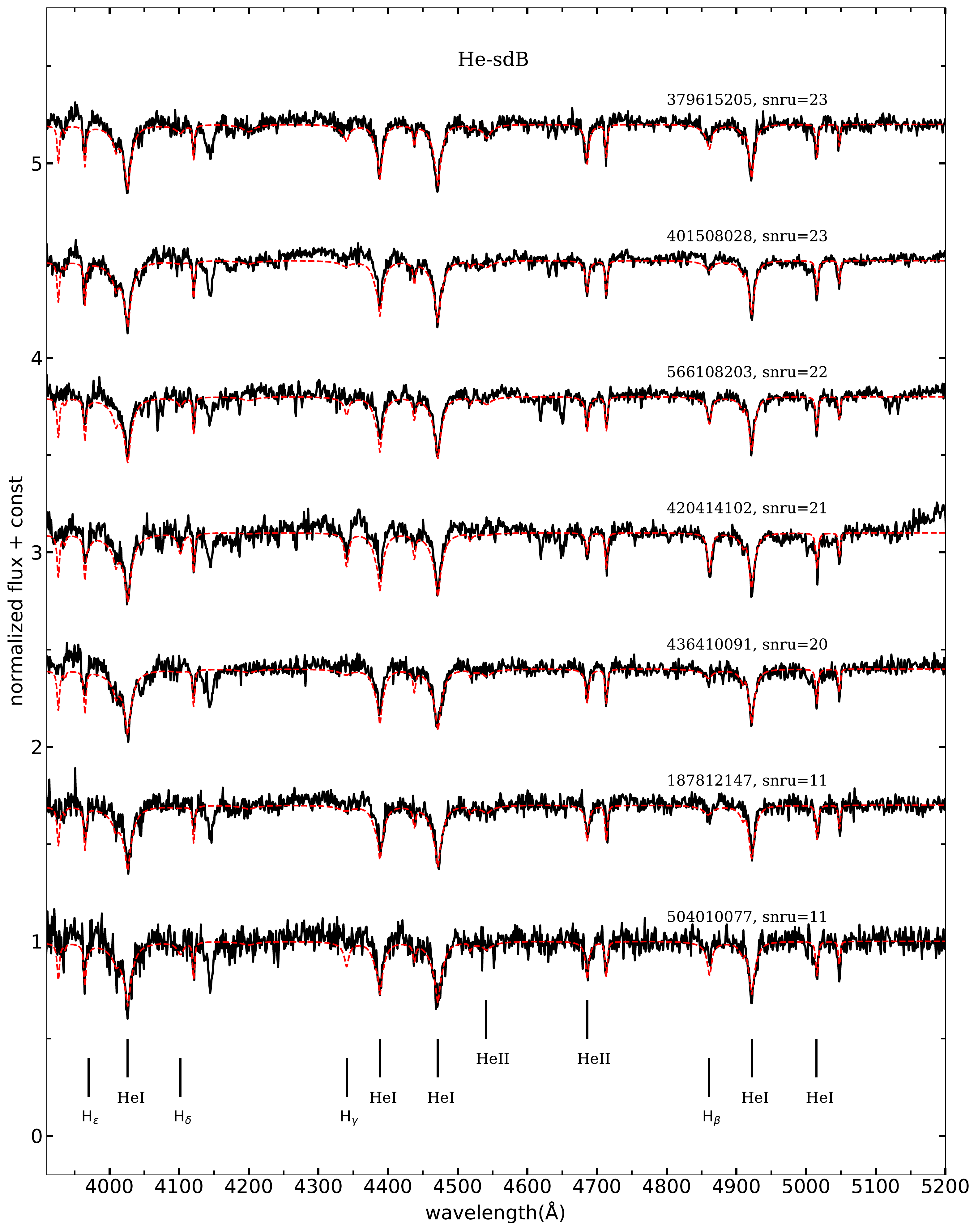} 
\centerline{(d) }
\end{minipage}%

\caption{Some fit examples for different types of hot subdwarf 
stars in our sample. The SNR in \textit{u} band decreases from top to bottom 
in each panel. The important H and He lines are marked at the bottom of each panel. } 
\end{figure}

\setcounter{figure}{2}
\begin{figure}
\centering

\begin{minipage}[c]{0.42\textwidth}
\includegraphics [width=75mm]{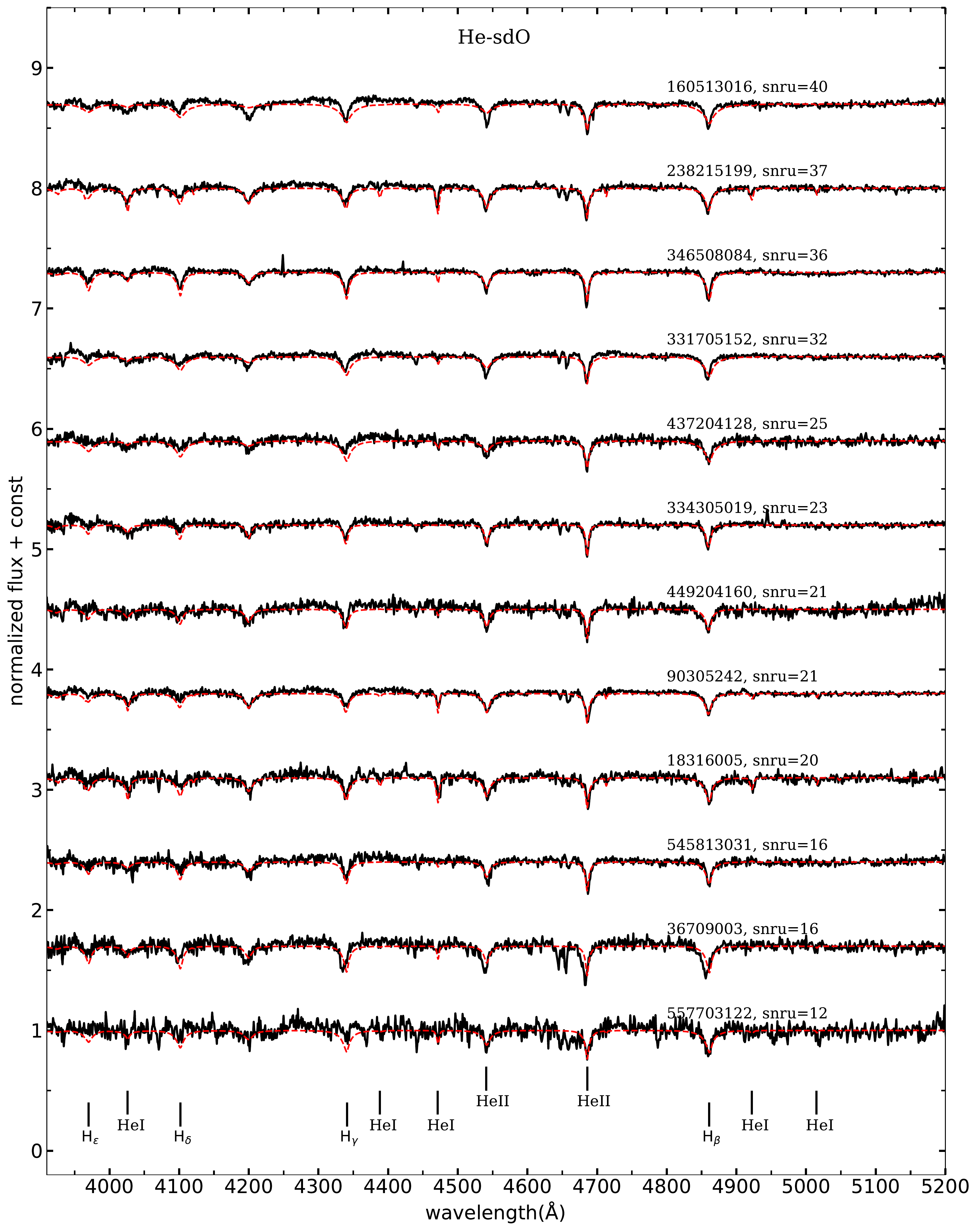} 
\centerline{(e) }
\end{minipage}%
\begin{minipage}[c]{0.42\textwidth}
\includegraphics [width=75mm]{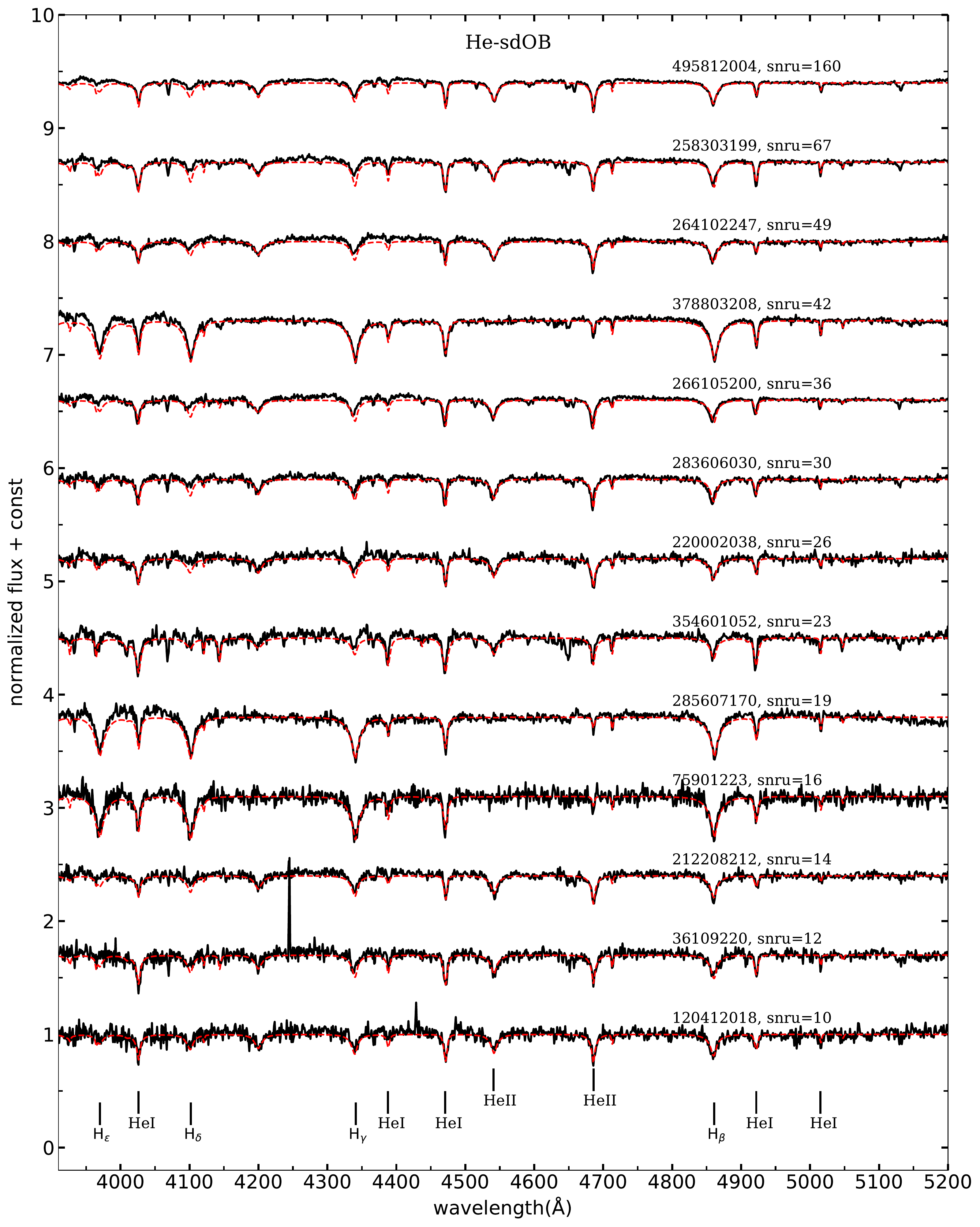} 
\centerline{(f) }
\end{minipage}%

\caption{Continued.} 
\end{figure}

We employed the spectral analysis tool, {\sc XTgrid} (N\'emeth et al. 2012, 2014), to analyze the observed 
LAMOST spectra as we did in Paper I. This program fits the observed data with synthetic spectra ({\sc Synspec} version 49; Lanz et al 2007) calculated from non-LTE model atmopsheres ({\sc Tlusty} version 204; Hubeny \& Lanz 2017). The best fitting model is searched for iteratively with a successive approximation method along the steepest-gradient of the $\chi^2$ field. The parameter uncertainties have been estimated by mapping the $\Delta\chi^2$ field until the 60 per cent confidence level at the given number of free parameters was reached.  
Standard uncertainties in {\sc XTgrid} are derived in one dimension, 
because parameter correlation calculations are extremely costly with 
non-LTE models and global, grid-less fitting. The global
minimization procedure is pursued until all gradients vanish 
and parameter correlations are minimal. Then each
parameter is changed until the 60 percent confidence level is reached to
find the corresponding error bars. Our 
procedure applies the degree of freedom corresponding to the number of
free parameters to set up confidence interval limits. 
Then a parabola is fitted to the data points obtained for a given
parameter, and the error bar is measured from the parabolic fit.  

We have upgraded {\sc XTgrid} with a new procedure to calculated two dimensional 
errors and uncover the correlation between $T_{\rm eff}$-$\log{g}$. 
Fig 2 shows an example of the $T_{\rm eff}$-$\log{g}$ correlation 
and error determination for a sdB star 
(LAMOST observation ID (obsid): 223907015). 
The color bar shows the chi-square variations with the
parameters. {\sc XTgrid} follows the chi-square gradients of 
all parameters to find the best fit for the observation.
The final model is represented here by the intersection 
of the black lines. During error calculations new models
are calculated radially outward from the best fit until 
the confidence limit is reached. The data points along the
lines represent the values that have been calculated from model atmospheres. 
The missing grid points have been
interpolated from their neighbors. The color coding is the 
interpolated and smoothed chi-square field around
the best fit. The contours show the confidence intervals for 60, 90 and 99\%. 
From these data asymmetric error bars are derived. 
Then, in an additional step, the asymmetric error bars are recalculated to symmetric errors, as
represented by the white error bars.

Fig 3 gives some examples of the best fitting models for different subdwarf types. In each panel, the black solid 
curves represent the normalized LAMOST spectra, while the red dashed curves represent 
the best-fitting synthetic spectra. The long integers in the right of panel 
are the LAMOST obsids. 
From Panel (a) to (f), the best fitting models for sdB, sdO, sdOB, He-sdB, 
He-sdO and He-sdOB are presented respectively. In each panel, the SNR of \textit{u} band decreases from top 
to bottom. By fitting the observed spectra, we obtained the atmospheric parameters 
for the 441 hot subdwarf candidates. As we did in Paper I, candidates with 
$T_{\rm eff}\ge$ 20\,000 K and $\log{g}\ge$ 5.0 $\mathrm{cm\ s^{-2}}$ are 
identified as our hot subdwarf stars, while candidates with 
$\log{g}<$ 5.0 $\mathrm{cm\ s^{-2}}$ and $\log{g}<$ 4.5 $\mathrm{cm\ s^{-2}}$ 
are considered as blue horizontal branch (BHB) stars and B type MS stars, 
respectively (N\'emeth et al. 2012).

We adopted the same classification 
scheme as in Paper I to classify our hot subdwarfs 
(see Paper I, Moehler et al. 1990, Geier et al. 2017a). 
Stars with dominant H Balmer lines, but no or weak He lines are 
classified as sdB stars. Stars with dominant H Balmer lines together 
with weak He I and He II lines are sdOB stars; Stars which present 
dominant H Balmer lines together with weak He II lines are sdO stars. 
Stars with dominant He II lines together with weak or no He I and H Balmer lines are He-sdO stars.
Stars that show dominant He I lines, but  weak He II and H Balmer lines are 
He-sdOB stars, while stars with dominant He I lines, but no He II and H Balmer lines are 
He-sdB stars \footnote{Actually, He-sdB stars  could present a wide 
range of He abundance and temperature 
 (Ahmad \& Jeffery 2006; Naslim et al. 2010), e.g., a type of intermediate 
 He-sdB (e.g., iHe-sdB, -1$<\mathrm{log}(n\mathrm{He}/n\mathrm{H}<$1.0)  
 and a type of extremely He-sdB (e.g., eHe-sdB, $\mathrm{log}(n\mathrm{He}/n\mathrm{H}>$1.0) 
 are present in Naslim et al. (2010).  
However, the He-sdB stars in our sample represent the hot subdwarfs with   
 the highest helium abundance (e.g., $\mathrm{log}(n\mathrm{He}/n\mathrm{H}>$1.0).  
Therefore,  one can easily make it clear that 
 the He-sdB stars in our study correspond to 
 eHe-sdB stars in Naslim et al. (2010), while our He-sdOB stars correspond to 
 iHe-sdB stars in Naslim et al. (2010).}.

\section{Results}
\begin{table*}
\tiny

 \begin{minipage}{160mm}
  \caption{Information on the 388 hot subdwarf stars identified in this study. From left to right, it gives 
  the right ascension (RA), declination (DEC), LAMOST\_obsid,  and Gaia source\_id. Next the $T_{\rm eff}$, $\log{g}$ and $\log(n{\rm He}/n{\rm H})$ are listed from the {\sc XTgrid} fits. Next, the SNR for the $u$ band,  the apparent magnitudes in the Gaia $G$ band and spectral classification are listed, respectively.  }
  \end{minipage}\\
  \centering
    \begin{tabularx}{16.0cm}{lllccccccccccccccX}
\hline\noalign{\smallskip}
RA\tablenotemark{a}  & DEC\tablenotemark{b}   &obsid\tablenotemark{c} & source\_id 
& $T_\mathrm{eff}$ &  $\mathrm{log}\ g$ & $\mathrm{log}(n\mathrm{He}/n\mathrm{H})$\tablenotemark{d}
&SNRU & $G$ & spclass  \\
 LAMOST &  LAMOST& LAMOST&Gaia  &(K)&($\mathrm{cm\ s^{-2}}$)& &&Gaia(mag) & \\
\hline\noalign{\smallskip}
0.065682$^{*}$ & 17.648047 & 472903166 & 2773760665114010880 & 36610$\pm$810 & 6.06$\pm$0.13 & -1.85$\pm$0.09 & 37 & 16.55 & sdOB \\
0.535283$^{*}$ & 19.987016 & 385805170 & 2846319155418023424 & 31560$\pm$450 & 5.58$\pm$0.06 & -3.00> & 15 & 15.60 & sdB \\
2.188996$^{*}$ & 12.288674 & 66613068 & 2766032919436757504 & 32230$\pm$330 & 6.00$\pm$0.08 & -2.26$\pm$0.13 & 15 & 14.62 & sdB \\
2.923521 & 19.340418 & 364410161 & 2797293512483062912 & 27310$\pm$170 & 5.45$\pm$0.03 & -2.84$\pm$0.08 & 49 & 14.74 & sdB \\
3.593379$^{*}$ & 28.615494 & 2915216 & 2859984160805522176 & 24180$\pm$80 & 5.44$\pm$0.04 & -3.00> & 45 & 12.65 & sdB \\
4.681331$^{*}$ & 1.023214 & 266708166 & 2547139876837313920 & 29090$\pm$150 & 5.47$\pm$0.05 & -3.32$\pm$0.07 & 44 & 14.87 & sdB \\
4.822343 & 40.502298 & 475310178 & 380384779299991424 & 26620$\pm$640 & 5.07$\pm$0.06 & -2.71$\pm$0.26 & 15 & 16.34 & sdB \\
5.353299 & 40.482537 & 90110033 & 380338531092183424 & 26580$\pm$520 & 5.34$\pm$0.04 & -2.41$\pm$0.12 & 19 & 15.51 & sdB \\
5.98014$^{*}$ & 42.151544 & 90103101 & 382086995098288896 & 30320$\pm$220 & 5.55$\pm$0.06 & -2.41$\pm$0.09 & 18 & 15.79 & sdB \\
6.36564$^{*}$ & 8.965084 & 248813198 & 2750667003920099200 & 29700$\pm$180 & 5.39$\pm$0.05 & -2.79$\pm$0.23 & 16 & 15.60 & sdB \\
6.439609 & 31.591527 & 284601234 & 2862287843823854848 & 26180$\pm$340 & 5.39$\pm$0.03 & -2.27$\pm$0.05 & 38 & 15.52 & sdB \\
6.528333 & 44.555064 & 382607056 & 382781542850500352 & 35770$\pm$330 & 5.75$\pm$0.07 & -1.63$\pm$0.09 & 13 & 17.25 & sdOB \\
6.949183$^{*}$ & 34.674052 & 75901223$^{\ddagger}$ & 365939429892433920 & 33010$\pm$470 & 5.72$\pm$0.04 & -0.76$\pm$0.04 & 16 & 15.77 & He-sdOB \\
7.287603$^{*}$ & 4.939857 & 182909073 & 2554910434747250944 & 33870$\pm$450 & 5.68$\pm$0.07 & -3.29> & 20 & 14.98 & sdB \\
9.522662 & 43.74734 & 472206190 & 387661209812972928 & 36180$\pm$640 & 5.85$\pm$0.10 & -1.46$\pm$0.15 & 12 & 16.15 & sdOB \\
12.949002 & 27.003799 & 157506158 & 2808769596378258048 & 31780$\pm$330 & 5.82$\pm$0.14 & -1.95$\pm$0.13 & 13 & 15.81 & sdB \\
16.574205 & 41.576104 & 182701066 & 374540943716955392 & 26090$\pm$910 & 5.26$\pm$0.07 & -2.18$\pm$0.12 & 24 & 15.96 & sdB \\
18.225359 & 32.633936 & 386014113 & 313700498585403776 & 44940$\pm$550 & 5.73$\pm$0.10 & 0.82$\pm$0.09 & 15 & 16.05 & He-sdOB \\
18.822416$^{*}$ & 14.471761 & 385201146 & 2590919234398332544 & 42160$\pm$5940 & 5.49$\pm$0.10 & -2.14> & 13 & 15.32 & sdO \\
18.976123$^{*}$ & 26.233403$^{\dagger}$ & 90305242 & 294818409307323904 & 55970$\pm$1350 & 5.68$\pm$0.10 & 0.58$\pm$0.13 & 21 & 14.42 & He-sdO \\
19.738333$^{*}$ & -0.429333 & 20615234$^{\ddagger}$ & 2533806305483837440 & 29060$\pm$60 & 5.53$\pm$0.01 & -3.15$\pm$0.15 & 16 & 14.82 & sdB \\
20.095589$^{*}$ & 39.849842 & 405010223 & 371711041304831616 & 28900$\pm$210 & 5.35$\pm$0.02 & -2.84$\pm$0.09 & 34 & 15.41 & sdB \\
20.91954$^{*}$ & 30.04227 & 386001132 & 309391787394140288 & 29260$\pm$340 & 5.34$\pm$0.05 & -1.97$\pm$0.09 & 11 & 16.50 & sdB \\
21.126275$^{*}$ & 6.816381 & 476409178 & 2566039072968204288 & 33890$\pm$350 & 5.66$\pm$0.04 & -1.78$\pm$0.08 & 15 & 16.44 & sdOB \\
21.587303 & 1.575993 & 354801221 & 2558457317524344704 & 49470$\pm$5840 & 5.92$\pm$0.14 & -2.24$\pm$0.27 & 13 & 17.33 & sdO \\
22.893792 & 32.623252 & 15209050 & 316170482737881728 & 73770$\pm$5630 & 5.34$\pm$0.13 & -1.36$\pm$0.12 & 14 & 15.30 & sdO \\
23.574204 & 26.388729 & 267314156 & 295620885292213248 & 28950$\pm$720 & 5.33$\pm$0.08 & -2.44$\pm$0.54 & 14 & 16.28 & sdB \\
26.036859 & 54.036332 & 468901095 & 408127652396870400 & 44900$\pm$520 & 5.79$\pm$0.42 & -2.34$\pm$0.10 & 40 & 16.24 & sdO \\
26.587053 & 45.69302 & 486404140 & 350797947892605952 & 29370$\pm$90 & 5.80$\pm$0.02 & -2.83$\pm$0.10 & 28 & 17.45 & sdB \\
29.296454$^{*}$ & 20.654967 & 378509232 & 97013488627139328 & 34630$\pm$740 & 5.83$\pm$0.14 & -1.54$\pm$0.07 & 21 & 15.39 & sdOB \\
29.942499$^{*}$ & 44.22208 & 191610229 & 346721341030064640 & 43970$\pm$1190 & 5.72$\pm$0.16 & 1.51$\pm$0.66 & 15 & 15.31 & He-sdOB \\
31.084296 & 29.459688 & 293313054 & 300019923940638336 & 29590$\pm$60 & 5.65$\pm$0.02 & -3.00> & 13 & 16.20 & sdB \\
31.265314 & 46.264437 & 306908165 & 352951036474046848 & 26900$\pm$100 & 5.25$\pm$0.03 & -1.60$\pm$0.03 & 120 & 11.52 & sdB \\
32.385219$^{*}$ & 43.12014 & 83912135 & 351633546665907200 & 26570$\pm$360 & 5.42$\pm$0.05 & -2.63$\pm$0.19 & 11 & 14.34 & sdB \\
32.725888$^{*}$ & 1.796436 & 290209126 & 2514184283536110720 & 42940$\pm$510 & 5.59$\pm$0.11 & 0.39$\pm$0.10 & 34 & 13.68 & He-sdOB \\
33.959493 & 29.345947 & 405907157 & 108002248353357312 & 24270$\pm$90 & 5.63$\pm$0.01 & -3.00> & 46 & 14.01 & sdB \\
34.935812$^{*}$ & 24.790013 & 379615205 & 102803104542235008 & 39860$\pm$680 & 5.42$\pm$0.37 & 2.03$\pm$0.04 & 23 & 16.38 & He-sdB \\
35.111382$^{*}$ & 24.324422 & 386615187 & 102583301000450688 & 31040$\pm$320 & 5.67$\pm$0.07 & -3.15> & 22 & 16.76 & sdB \\
35.358014 & 8.988819 & 266809027 & 23700286669971584 & 40690$\pm$20 & 5.53$\pm$0.42 & 1.53$\pm$0.05 & 20 & 17.00 & He-sdOB \\
36.005261$^{*}$ & 21.947266 & 379608221 & 99836961472298752 & 37010$\pm$370 & 5.86$\pm$0.06 & -1.01$\pm$0.06 & 40 & 15.68 & sdOB \\
36.301804$^{*}$ & 23.806172$^{\dagger}$ & 379609016 & 101826669497182720 & 52640$\pm$1380 & 5.46$\pm$0.08 & -1.92$\pm$0.05 & 90 & 13.84 & sdO \\
38.225357 & 37.072618 & 176806221 & 333636534183932416 & 35590$\pm$930 & 5.66$\pm$0.12 & 0.28$\pm$0.76 & 17 & 15.85 & He-sdOB \\
40.018332$^{*}$ & 3.928428 & 160513016 & 2503706418759709184 & 65870$\pm$1270 & 5.19$\pm$0.16 & 0.19$\pm$0.08 & 40 & 13.98 & He-sdO \\
41.412008$^{*}$ & 13.432122 & 298408084 & 31918414533148928 & 32120$\pm$130 & 5.34$\pm$0.01 & -1.53$\pm$0.02 & 27 & 13.21 & sdB \\
43.689377 & 9.275102 & 369315189 & 20948346504556160 & 29370$\pm$270 & 5.65$\pm$0.05 & -3.00> & 12 & 17.71 & sdB \\
45.303592$^{*}$ & 18.681715 & 387107175 & 35928500943041152 & 30240$\pm$150 & 5.43$\pm$0.03 & -3.00> & 38 & 15.42 & sdB \\
45.548188$^{*}$ & 37.185967 & 163303129 & 142086971376178688 & 44530$\pm$1700 & 5.63$\pm$0.11 & -2.54$\pm$0.23 & 28 & 15.50 & sdO \\
45.650063 & 28.271376 & 249704157 & 116691791828603136 & 37070$\pm$230 & 5.87$\pm$0.07 & -1.59$\pm$0.05 & 38 & 16.27 & sdOB \\
48.4049$^{*}$ & 15.105965 & 406503078 & 31009771252186752 & 42110$\pm$180 & 5.63$\pm$0.02 & 1.32$\pm$0.05 & 52 & 15.62 & He-sdOB \\
50.255387 & 18.955313 & 417602042 & 56252977678408192 & 27360$\pm$540 & 5.40$\pm$0.05 & -2.52$\pm$0.16 & 11 & 16.36 & sdB \\
50.388316 & 8.192122 & 414808170 & 11015044926034816 & 33610$\pm$510 & 5.69$\pm$0.07 & -1.74$\pm$0.05 & 23 & 15.94 & sdOB \\
50.411157$^{*}$ & 5.644444 & 8315230$^{\ddagger}$ & 9132341717393792 & 31550$\pm$220 & 5.84$\pm$0.05 & -2.06$\pm$0.03 & 28 & 14.99 & sdB \\
55.122076 & 24.499152 & 474303164 & 68334308365794304 & 24940$\pm$420 & 5.66$\pm$0.07 & -3.00> & 11 & 16.55 & sdB \\
55.654778 & 20.022511 & 249804028 & 63139734399542528 & 44340$\pm$780 & 5.96$\pm$0.08 & 0.64$\pm$0.06 & 16 & 16.21 & He-sdOB \\
58.70092 & 35.255844 & 301309205 & 218951870771462144 & 77920$\pm$4860 & 7.06$\pm$0.16 & -1.28$\pm$0.23 & 24 & 16.84 & sdO \\
59.293668 & 28.828236 & 269903217 & 166876629257005696 & 33470$\pm$630 & 5.31$\pm$0.10 & -2.73$\pm$0.26 & 13 & 17.43 & sdB \\
61.095407 & 26.131827 & 170611197 & 162522700650966400 & 30350$\pm$180 & 5.75$\pm$0.06 & -3.34$\pm$0.22 & 15 & 15.72 & sdB \\
62.768174 & 34.310777 & 200512196 & 170774775937432832 & 26840$\pm$480 & 5.43$\pm$0.05 & -3.04$\pm$0.08 & 19 & 16.38 & sdB \\
62.954352$^{*}$ & 15.383123 & 254112160 & 3311822554365339776 & 37490$\pm$520 & 5.59$\pm$0.04 & -3.23$\pm$0.15 & 56 & 14.67 & sdO \\
64.404255 & 30.127843 & 504604071 & 165712109004835712 & 31570$\pm$190 & 5.60$\pm$0.04 & -2.09$\pm$0.06 & 21 & 17.06 & sdB \\
66.426054 & -6.159812 & 386702207 & 3199598460537019008 & 42260$\pm$580 & 5.45$\pm$0.08 & 1.64$\pm$0.30 & 20 & 16.69 & He-sdOB \\
71.786429 & 21.890949 & 197202040 & 3412738686500055296 & 30430$\pm$300 & 5.39$\pm$0.08 & -2.70$\pm$0.15 & 16 & 17.69 & sdB \\
72.615315 & 15.986111 & 402814084 & 3405118997904684032 & 37480$\pm$270 & 5.73$\pm$0.07 & -0.90$\pm$0.08 & 16 & 16.64 & He-sdOB \\
74.313203 & 43.055659 & 422313143 & 205078336132788352 & 27680$\pm$540 & 5.48$\pm$0.08 & -2.87$\pm$0.27 & 14 & 16.03 & sdB \\
75.717784 & 16.44665 & 266202113 & 3393408962846662656 & 44970$\pm$1750 & 5.16$\pm$0.19 & -3.03$\pm$0.37 & 28 & 15.61 & sdO \\
81.110577 & 15.08733 & 498205197 & 3390739005017699968 & 35140$\pm$470 & 5.45$\pm$0.06 & -3.11$\pm$0.21 & 12 & 16.93 & sdB \\
81.286539 & 17.953808 & 374416146 & 3400347392320018176 & 34810$\pm$540 & 5.68$\pm$0.09 & -2.30$\pm$0.09 & 30 & 15.67 & sdOB \\
81.961523 & 16.909214 & 374415024 & 3397199529183763072 & 45490$\pm$730 & 5.40$\pm$0.09 & 0.62$\pm$0.13 & 40 & 15.66 & He-sdOB \\
82.391388 & 4.122033 & 506809203 & 3236488827993523584 & 33980$\pm$2120 & 6.00$\pm$0.12 & -0.93$\pm$0.11 & 12 & 16.53 & He-sdOB \\
\hline\noalign{\smallskip} 
  \end{tabularx}
  \tablenotetext{a}{Stars labeled with $\ast$ also appear in the hot subdwarf catalog of Geier et al. (2017a).}  \tablenotetext{b}{Stars labeled with $\dagger$ also appear in N\'emeth et al. (2012).} 
 \tablenotetext{c}{Stars labeled with $\ddagger$ also appear in Luo et al. (2016).} 
 \tablenotetext{d}{"$>$" denotes an upper limit of $\mathrm{log}(n\mathrm{He}/n\mathrm{H})$ for the object.} 
\end{table*}

\setcounter{table}{0}
\begin{table*}
\tiny

 \begin{minipage}{160mm}
  \caption{Continued.}
  \end{minipage}\\
  \centering
    \begin{tabularx}{16.0cm}{lllccccccccccccccX}
\hline\noalign{\smallskip}
RA\tablenotemark{a}  & DEC\tablenotemark{b}   &obsid\tablenotemark{c} & source\_id 
& $T_\mathrm{eff}$ &  $\mathrm{log}\ g$ & $\mathrm{log}(n\mathrm{He}/n\mathrm{H})$\tablenotemark{d}
&SNRU & $G$  & spclass  \\
 LAMOST &  LAMOST& LAMOST&Gaia  &(K)&($\mathrm{cm\ s^{-2}}$)& && Gaia(mag)& \\
\hline\noalign{\smallskip}
83.163266 & 43.638849 & 499503037 & 195587076882426880 & 30010$\pm$220 & 5.58$\pm$0.10 & -2.74$\pm$0.18 & 12 & 16.36 & sdB \\
83.348132 & 21.438994 & 198702152 & 3403657609512772096 & 48580$\pm$1160 & 5.51$\pm$0.06 & -3.00> & 16 & 16.90 & sdO \\
85.570863 & 12.830711 & 506005080 & 3340527779312131584 & 31100$\pm$390 & 5.59$\pm$0.05 & -2.81> & 11 & 16.84 & sdB \\
86.096935 & 38.297225 & 381414035 & 189831438452843008 & 28200$\pm$380 & 5.16$\pm$0.06 & -2.84$\pm$0.24 & 11 & 16.06 & sdB \\
86.672165 & 32.32837 & 174308109 & 3448218449263806080 & 33050$\pm$460 & 5.90$\pm$0.13 & -2.05$\pm$0.11 & 10 & 17.15 & sdOB \\
86.798328 & 17.15543 & 420414102 & 3396486324095591040 & 36050$\pm$1680 & 5.53$\pm$0.08 & 1.52$\pm$0.04 & 21 & 16.40 & He-sdB \\
87.875844 & 38.478464 & 381403073 & 3457951566709510912 & 33180$\pm$530 & 5.80$\pm$0.08 & -2.26$\pm$0.10 & 20 & 16.12 & sdB \\
87.980862 & 13.398042 & 404506102 & 3346423532459289088 & 32240$\pm$200 & 5.68$\pm$0.05 & -1.70$\pm$0.05 & 24 & 16.05 & sdOB \\
90.000993 & 11.476859 & 401508028 & 3342199204486994048 & 38530$\pm$470 & 5.57$\pm$0.09 & 2.69$\pm$0.04 & 23 & 16.28 & He-sdB \\
93.024484 & 47.523367 & 546803066 & 969542451362654976 & 47330$\pm$830 & 5.05$\pm$0.10 & 0.49$\pm$0.06 & 35 & 16.22 & He-sdOB \\
93.355911$^{*}$ & 34.34843$^{\dagger}$ & 177305180 & 3452237404778065152 & 34880$\pm$430 & 5.78$\pm$0.04 & -1.39$\pm$0.04 & 50 & 13.67 & sdOB \\
93.506639$^{*}$ & 33.491003 & 187306026 & 3440046359151100032 & 52700$\pm$2130 & 5.30$\pm$0.15 & -3.32$\pm$0.34 & 24 & 16.70 & sdO \\
93.633112 & -7.131204 & 211406205 & 3007619194841792640 & 29540$\pm$140 & 5.43$\pm$0.05 & -3.33$\pm$0.14 & 74 & 13.94 & sdB \\
93.792659$^{*}$ & 35.854813 & 185915228 & 3452919960980353920 & 28610$\pm$280 & 5.68$\pm$0.04 & -2.58$\pm$0.06 & 28 & 16.45 & sdB \\
93.949651$^{*}$ & 34.780417 & 220504128 & 3452646281368418304 & 22640$\pm$550 & 5.71$\pm$0.07 & -3.00> & 25 & 16.06 & sdB \\
95.688418 & 51.846152 & 551803036 & 995383482875359744 & 32360$\pm$590 & 5.61$\pm$0.10 & -1.94$\pm$0.09 & 24 & 16.48 & sdOB \\
95.878258 & 33.022396 & 185907227 & 3439445854003662592 & 38500$\pm$1070 & 5.35$\pm$0.09 & -3.00> & 30 & 16.19 & sdO \\
96.176952 & 28.496275 & 332303054 & 3433651805682659200 & 32930$\pm$710 & 5.87$\pm$0.09 & -1.80$\pm$0.12 & 16 & 17.23 & sdOB \\
96.799818 & 23.152877 & 301408154 & 3377315754750660992 & 39790$\pm$390 & 6.19$\pm$0.07 & -1.72$\pm$0.09 & 11 & 15.94 & sdOB \\
97.454305 & 22.414013 & 274411112 & 3376468958998845312 & 28110$\pm$280 & 5.57$\pm$0.05 & -2.65$\pm$0.17 & 14 & 15.43 & sdB \\
97.865182 & 50.14843 & 551801179 & 968154867688361856 & 25350$\pm$720 & 5.34$\pm$0.07 & -3.08$\pm$0.27 & 22 & 16.60 & sdB \\
98.397232 & 23.148759 & 220002038 & 3382340316731002368 & 46790$\pm$440 & 5.61$\pm$0.06 & 0.75$\pm$0.09 & 26 & 15.51 & He-sdOB \\
98.719332 & 37.186158 & 302804228 & 943173001432279040 & 45470$\pm$960 & 5.20$\pm$0.10 & 0.56$\pm$0.10 & 17 & 17.50 & He-sdOB \\
99.208728 & 29.323616 & 332312170 & 3434651369127681664 & 39060$\pm$410 & 5.56$\pm$0.09 & -0.11$\pm$0.08 & 14 & 17.14 & He-sdOB \\
99.308942 & 42.381702 & 317915162 & 958032042088084992 & 34870$\pm$450 & 5.97$\pm$0.07 & -1.34$\pm$0.07 & 11 & 17.25 & sdOB \\
99.660973 & 42.948122 & 475702062 & 964071899977558784 & 29760$\pm$270 & 5.37$\pm$0.03 & -2.86$\pm$0.17 & 20 & 16.17 & sdB \\
100.62136$^{*}$ & 37.070349 & 206006056 & 943819308110646144 & 34830$\pm$330 & 5.82$\pm$0.10 & -1.56$\pm$0.05 & 40 & 15.09 & sdOB \\
101.28447 & 21.304168 & 319005070 & 3378155777335873664 & 27260$\pm$260 & 5.30$\pm$0.07 & -2.52$\pm$0.24 & 15 & 16.16 & sdB \\
101.79684 & 13.370346 & 387309021 & 3352595366104656384 & 35580$\pm$250 & 6.03$\pm$0.04 & -1.26$\pm$0.07 & 12 & 17.39 & sdOB \\
101.89525 & 13.836488 & 387309030 & 3352835983053282304 & 31770$\pm$220 & 6.06$\pm$0.04 & -2.46$\pm$0.18 & 18 & 16.83 & sdB \\
102.31712$^{*}$ & 38.572141 & 296103019 & 944390774983674496 & 33450$\pm$140 & 5.72$\pm$0.03 & -1.96$\pm$0.06 & 41 & 15.52 & sdOB \\
102.6132 & 16.662794 & 282306209 & 3358122782854938112 & 28870$\pm$350 & 5.65$\pm$0.05 & -3.00> & 13 & 16.32 & sdB \\
102.83762 & 52.007091 & 546308112 & 993122779593859584 & 30300$\pm$420 & 5.82$\pm$0.09 & -2.38$\pm$0.18 & 13 & 16.30 & sdB \\
103.216529$^{*}$ & 29.006594 & 15710159$^{\ddagger}$ & 887620515740571264 & 32270$\pm$380 & 5.74$\pm$0.06 & -1.91$\pm$0.10 & 26 & 14.79 & sdB \\
103.23504 & 12.049997 & 387306122 & 3351464724552173952 & 46200$\pm$900 & 5.70$\pm$0.10 & 0.01$\pm$0.07 & 11 & 17.13 & He-sdOB \\
103.80732 & 15.948556 & 422403108 & 3355060020198388480 & 29400$\pm$420 & 5.40$\pm$0.13 & -3.29$\pm$0.40 & 12 & 16.45 & sdB \\
104.47756 & 12.242259 & 392004058 & 3160786559070298880 & 50480$\pm$780 & 5.50$\pm$0.08 & -1.77$\pm$0.15 & 44 & 15.69 & sdO \\
105.11577 & 23.132122 & 298605046 & 3380129503787793920 & 29250$\pm$200 & 5.48$\pm$0.06 & -2.60$\pm$0.10 & 12 & 17.05 & sdB \\
105.2623 & 9.693231 & 507814235 & 3157919887443293952 & 23300$\pm$610 & 5.56$\pm$0.05 & -3.35$\pm$0.30 & 11 & 17.05 & sdB \\
105.34676$^{*}$ & 53.493837$^{\dagger}$ & 545813031 & 981727582818800128 & 73480$\pm$4480 & 5.07$\pm$0.12 & 0.85$\pm$0.65 & 16 & 14.93 & He-sdO \\
105.449625$^{*}$ & 28.568161 & 21003121$^{\ddagger}$ & 884393900085830144 & 26710$\pm$500 & 5.40$\pm$0.09 & -1.92$\pm$0.11 & 18 & 14.90 & sdB \\
105.52333 & 13.673537 & 422401101 & 3353521803431070464 & 31610$\pm$160 & 5.76$\pm$0.05 & -1.81$\pm$0.07 & 12 & 17.12 & sdB \\
105.56337$^{*}$ & 12.29816 & 392009130 & 3160836861727862400 & 27110$\pm$410 & 5.41$\pm$0.04 & -3.18$\pm$0.10 & 46 & 14.74 & sdB \\
105.66879 & 11.794911 & 392008093 & 3160522848078571008 & 42690$\pm$520 & 5.62$\pm$0.07 & 0.76$\pm$0.08 & 28 & 15.79 & He-sdOB \\
105.73661 & 17.391982 & 372812146 & 3361273433424723712 & 29210$\pm$170 & 5.58$\pm$0.03 & -2.13$\pm$0.07 & 20 & 15.78 & sdB \\
105.77822 & 19.537799 & 174905039 & 3364990572999194880 & 37730$\pm$760 & 5.40$\pm$0.08 & -3.21$\pm$0.32 & 16 & 15.93 & sdB \\
106.64127 & 9.090862 & 507803187 & 3154729899269006336 & 28260$\pm$240 & 5.37$\pm$0.02 & -2.85$\pm$0.16 & 25 & 16.35 & sdB \\
107.11835 & 44.426357 & 436410091 & 953141929764017408 & 38760$\pm$750 & 5.98$\pm$0.48 & 3.35$\pm$0.09 & 20 & 16.30 & He-sdB \\
108.15275 & 13.195084 & 389404129 & 3161149775864814720 & 38280$\pm$900 & 5.97$\pm$0.07 & -1.43$\pm$0.07 & 13 & 17.24 & sdOB \\
108.22018 & 42.643905 & 316516061 & 949617548319629824 & 47000$\pm$1340 & 5.71$\pm$0.04 & -3.42$\pm$0.48 & 20 & 16.15 & sdO \\
108.27216 & 18.417765 & 267011180 & 3363070035424348800 & 27180$\pm$460 & 5.80$\pm$0.02 & -3.00> & 29 & 15.85 & sdB \\
108.31042 & 17.582952 & 267015014 & 3360729067791609216 & 29520$\pm$460 & 5.60$\pm$0.03 & -2.47$\pm$0.05 & 25 & 16.00 & sdB \\
108.33188 & 38.127736 & 187812147 & 946428174325145856 & 39780$\pm$930 & 5.58$\pm$0.12 & 3.65$\pm$0.06 & 11 & 16.69 & He-sdB \\
108.37492 & 19.895661 & 175006165 & 3363513924586196224 & 27480$\pm$110 & 5.33$\pm$0.15 & -2.56$\pm$0.12 & 15 & 16.88 & sdB \\
108.6558 & 26.786667 & 281105033 & 883333730358019584 & 24650$\pm$290 & 5.07$\pm$0.12 & -1.92$\pm$0.31 & 19 & 15.83 & sdB \\
109.71113 & 12.844212 & 389406022 & 3166142933042903424 & 26120$\pm$580 & 5.54$\pm$0.10 & -2.42$\pm$0.14 & 11 & 17.23 & sdB \\
109.73439 & 10.44399 & 446310014 & 3156573539750882944 & 31080$\pm$380 & 5.47$\pm$0.09 & 0.41$\pm$0.07 & 35 & 15.79 & He-sdOB \\
109.84141 & 5.16679 & 492414173 & 3140714527629703552 & 35180$\pm$260 & 5.79$\pm$0.12 & -1.56$\pm$0.04 & 24 & 16.39 & sdOB \\
110.995849 & 4.499076 & 492404174 & 3139698956841267456 & 29850$\pm$60 & 5.25$\pm$0.04 & -2.71$\pm$0.09 & 38 & 15.40 & sdB \\
111.42846 & 52.723794 & 448315232 & 986474685615378816 & 25000$\pm$170 & 5.50$\pm$0.02 & -3.00> & 56 & 15.31 & sdB \\
112.103$^{*}$ & 41.831571 & 320613193 & 948767385313600128 & 38230$\pm$320 & 5.71$\pm$0.06 & -0.11$\pm$0.07 & 13 & 15.73 & He-sdOB \\
112.74626 & 33.618403 & 320101130 & 893467615497884544 & 35750$\pm$290 & 5.80$\pm$0.38 & -2.55> & 20 & 17.19 & sdB \\
112.816215 & 3.47858 & 492406014 & 3136412997623472512 & 24850$\pm$290 & 5.09$\pm$0.04 & -2.44$\pm$0.12 & 13 & 16.57 & sdB \\
112.88153$^{*}$ & 20.756841 & 270109111 & 864917013671899648 & 50930$\pm$2040 & 5.50$\pm$0.09 & -1.58$\pm$0.11 & 28 & 15.64 & sdO \\
113.065169 & 6.203372 & 492412181 & 3141525486172958336 & 28910$\pm$170 & 5.54$\pm$0.08 & -3.08$\pm$0.50 & 21 & 15.93 & sdB \\
113.08392$^{*}$ & 27.069038 & 21602071 & 872322087046515840 & 36640$\pm$890 & 5.91$\pm$0.10 & 0.48$\pm$0.04 & 10 & 16.64 & He-sdOB \\
113.69226 & 34.355805 & 75815079 & 893594510306886784 & 26400$\pm$480 & 5.25$\pm$0.04 & -2.45$\pm$0.08 & 24 & 15.46 & sdB \\
114.30941 & 25.098511 & 228001126 & 868529562204066688 & 26760$\pm$410 & 5.29$\pm$0.06 & -2.34$\pm$0.12 & 11 & 15.52 & sdB \\
114.63907 & 51.47903 & 448308129 & 982729822026188928 & 26880$\pm$400 & 5.45$\pm$0.04 & -3.26$\pm$0.32 & 11 & 15.63 & sdB \\
114.73744$^{*}$ & 40.328337 & 331705152 & 923580185261224576 & 105770$\pm$970 & 5.24$\pm$0.03 & 0.36$\pm$0.04 & 32 & 16.45 & He-sdO \\
115.01244 & 39.775833 & 334305019 & 923342347152856576 & 89120$\pm$12800 & 5.13$\pm$0.05 & 0.34$\pm$0.07 & 23 & 15.54 & He-sdO \\
115.09819$^{*}$ & 20.826954 & 176203116 & 673058556816796288 & 33560$\pm$140 & 5.76$\pm$0.02 & -1.64$\pm$0.04 & 63 & 15.54 & sdOB \\
115.10468$^{*}$ & 15.8653 & 499603225 & 3165813590653493760 & 32430$\pm$660 & 6.21$\pm$0.12 & -3.00> & 11 & 17.07 & sdB \\
115.6535 & 13.175169 & 407404024 & 3163997712841285504 & 28330$\pm$410 & 5.27$\pm$0.04 & -2.72$\pm$0.17 & 14 & 17.55 & sdB \\
116.13832 & 59.994095 & 448915044 & 1086079547054989184 & 28470$\pm$570 & 5.64$\pm$0.07 & -2.59$\pm$0.18 & 11 & 17.02 & sdB \\
116.5139$^{*}$ & 33.743557 & 273608236 & 882125367078828672 & 26480$\pm$240 & 5.22$\pm$0.03 & -2.43$\pm$0.07 & 49 & 14.99 & sdB \\
116.55484$^{*}$ & 33.552105 & 120412018$^{\ddagger}$ & 882087811884800896 & 46150$\pm$560 & 5.51$\pm$0.10 & 0.67$\pm$0.09 & 10 & 15.85 & He-sdOB \\
117.04707$^{*}$ & 43.877754 & 403501067 & 925982721247295488 & 27430$\pm$530 & 5.54$\pm$0.06 & -1.75$\pm$0.08 & 13 & 16.99 & sdB \\
\hline\noalign{\smallskip} 
  \end{tabularx}
\end{table*}

\setcounter{table}{0}
\begin{table*}
\tiny

 \begin{minipage}{160mm}
  \caption{Continued.}
  \end{minipage}\\
  \centering
    \begin{tabularx}{16.0cm}{lllccccccccccccccX}
\hline\noalign{\smallskip}
RA\tablenotemark{a}  & DEC\tablenotemark{b}   &obsid\tablenotemark{c} & source\_id 
& $T_\mathrm{eff}$ &  $\mathrm{log}\ g$ & $\mathrm{log}(n\mathrm{He}/n\mathrm{H})$\tablenotemark{d}
&SNRU & $G$ & spclass  \\
 LAMOST &  LAMOST& LAMOST&Gaia  &(K)&($\mathrm{cm\ s^{-2}}$)& &&Gaia(mag) & \\
\hline\noalign{\smallskip}
117.04874$^{*}$ & 13.730365 & 417207114 & 3164383748798177792 & 22940$\pm$930 & 5.36$\pm$0.16 & -2.68$\pm$0.34 & 13 & 15.66 & sdB \\
117.21696$^{*}$ & 45.817713 & 405304165 & 927227948229851392 & 27690$\pm$520 & 5.74$\pm$0.07 & -2.41$\pm$0.15 & 15 & 16.07 & sdB \\
117.26186$^{*}$ & 12.803115 & 407409236 & 3152255890729006848 & 37580$\pm$840 & 5.53$\pm$0.09 & -3.26$\pm$0.37 & 18 & 16.68 & sdB \\
118.1429$^{*}$ & 16.268133$^{\dagger}$ & 414013194 & 666988874675710464 & 50420$\pm$710 & 5.67$\pm$0.10 & 0.62$\pm$0.13 & 13 & 14.46 & He-sdOB \\
118.47597$^{*}$ & 25.519389 & 173809084 & 873767017124252544 & 33540$\pm$740 & 5.86$\pm$0.04 & -1.87$\pm$0.12 & 30 & 15.68 & sdOB \\
119.29932$^{*}$ & 13.752347 & 392702198 & 654222239006166656 & 36110$\pm$900 & 5.99$\pm$0.16 & -1.54$\pm$0.17 & 18 & 15.93 & sdOB \\
119.845827 & 16.767125 & 18216091 & 667308798199419776 & 38880$\pm$900 & 5.17$\pm$0.07 & -2.70$\pm$0.16 & 25 & 14.42 & sdO \\
119.93739$^{*}$ & 37.250873 & 285607170 & 907291264093243520 & 33980$\pm$390 & 5.82$\pm$0.07 & -0.91$\pm$0.04 & 19 & 16.30 & He-sdOB \\
120.63138$^{*}$ & 39.374639 & 279209235 & 921064055980216320 & 35770$\pm$230 & 5.47$\pm$0.05 & -1.34$\pm$0.07 & 47 & 15.19 & sdOB \\
120.86637$^{*}$ & 34.361297 & 78615208 & 905650346067635968 & 39300$\pm$410 & 5.43$\pm$0.04 & -3.09$\pm$0.20 & 26 & 15.06 & sdO \\
121.61706$^{*}$ & 32.516504 & 78608160$^{\ddagger}$ & 902154929884483456 & 32930$\pm$240 & 5.65$\pm$0.07 & -1.30$\pm$0.06 & 16 & 15.39 & sdB \\
121.61938$^{*}$ & 24.349293 & 103802062 & 680914734899282560 & 28190$\pm$260 & 5.50$\pm$0.03 & -2.49$\pm$0.06 & 18 & 15.00 & sdB \\
121.99274 & 27.409538 & 2001056 & 683766356962547840 & 38780$\pm$630 & 5.39$\pm$0.11 & -3.06$\pm$0.23 & 55 & 14.11 & sdO \\
122.14079$^{*}$ & 18.039414 & 283606030 & 668963357040208512 & 45980$\pm$610 & 5.37$\pm$0.10 & 0.67$\pm$0.06 & 30 & 16.64 & He-sdOB \\
122.34152$^{*}$ & 16.144793 & 392712156 & 656428099850193664 & 37730$\pm$1080 & 5.54$\pm$0.12 & -3.00> & 23 & 15.74 & sdB \\
123.02008$^{*}$ & 13.867944 & 388802203$^{\ddagger}$ & 653631869983111552 & 27630$\pm$460 & 5.27$\pm$0.08 & -2.74$\pm$0.17 & 15 & 17.29 & sdB \\
123.42883$^{*}$ & 27.842981 & 404612236 & 684061777696508544 & 27430$\pm$220 & 5.52$\pm$0.04 & -2.52$\pm$0.09 & 19 & 16.32 & sdB \\
123.88002$^{*}$ & 24.872947 & 404606240 & 682305655764689408 & 35360$\pm$670 & 5.55$\pm$0.10 & -3.00> & 29 & 16.36 & sdB \\
124.032567$^{*}$ & 48.064086 & 198809068 & 931348922106072704 & 24740$\pm$610 & 5.50$\pm$0.04 & -3.08$\pm$0.11 & 41 & 15.18 & sdB \\
124.82385$^{*}$ & 42.560752 & 318008007 & 915859960070594176 & 32880$\pm$180 & 6.10$\pm$0.04 & -2.18$\pm$0.05 & 14 & 16.24 & sdOB \\
125.60942$^{*}$ & 39.68863 & 105910059$^{\ddagger}$ & 908526118730334592 & 31510$\pm$470 & 5.80$\pm$0.10 & -2.28$\pm$0.11 & 11 & 16.95 & sdB \\
125.917117 & 8.327003 & 334913110 & 599294211494840704 & 28110$\pm$240 & 5.75$\pm$0.04 & -2.48$\pm$0.16 & 12 & 14.70 & sdB \\
127.5257$^{*}$ & 47.864096 & 302901126 & 930769616916994048 & 27030$\pm$980 & 5.57$\pm$0.09 & -3.43$\pm$0.52 & 19 & 16.06 & sdB \\
129.016533$^{*}$ & 15.870972 & 43203186$^{\ddagger}$ & 657873717121550976 & 26790$\pm$490 & 5.56$\pm$0.05 & -2.49$\pm$0.08 & 14 & 15.40 & sdB \\
129.050164$^{*}$ & 19.298929 & 431714232 & 662771594747150976 & 35310$\pm$470 & 5.49$\pm$0.07 & -1.69$\pm$0.10 & 12 & 15.71 & sdOB \\
129.824905 & 10.631478 & 420802086 & 601444306482580992 & 27030$\pm$280 & 5.52$\pm$0.03 & -2.73$\pm$0.07 & 47 & 15.18 & sdB \\
130.870297$^{*}$ & 8.410645 & 503814238 & 597102541222525056 & 25720$\pm$430 & 5.64$\pm$0.05 & -3.00> & 19 & 16.45 & sdB \\
131.499846$^{*}$ & -2.658236 & 302304166 & 3072185163322482816 & 32540$\pm$780 & 5.86$\pm$0.09 & -2.37$\pm$0.05 & 21 & 15.72 & sdB \\
131.82875$^{*}$ & 23.0086 & 316011216 & 689367475352913280 & 60270$\pm$1140 & 5.41$\pm$0.06 & -1.61$\pm$0.07 & 36 & 15.28 & sdO \\
132.677077$^{*}$ & 2.76284 & 449404224 & 577999175230073728 & 27540$\pm$150 & 5.33$\pm$0.03 & -2.50$\pm$0.09 & 30 & 15.61 & sdB \\
135.7186$^{*}$ & 32.221134 & 205006060 & 712415021480712448 & 36030$\pm$1120 & 5.78$\pm$0.30 & -1.89$\pm$0.77 & 14 & 18.16 & sdOB \\
136.199038$^{*}$ & 31.547997 & 28303115$^{\ddagger}$ & 711773838699959296 & 38250$\pm$420 & 5.62$\pm$0.06 & -0.73$\pm$0.06 & 20 & 15.12 & He-sdOB \\
136.272592$^{*}$ & 5.550386 & 424616225 & 580372234855004160 & 41170$\pm$20 & 5.60$\pm$0.01 & 1.68$\pm$0.12 & 22 & 14.06 & He-sdOB \\
136.783847$^{*}$ & -3.103886 & 495812004 & 5762960596350664832 & 48630$\pm$1150 & 5.47$\pm$0.05 & 0.75$\pm$0.07 & 160 & 11.88 & He-sdOB \\
137.58958$^{*}$ & 59.50939 & 209716167 & 1039246845702047360 & 73690$\pm$7550 & 6.97$\pm$0.32 & -1.74$\pm$0.24 & 12 & 15.34 & sdO \\
138.536147$^{*}$ & 3.967868 & 424604230$^{\ddagger}$ & 578986163009656320 & 29420$\pm$370 & 5.52$\pm$0.07 & -2.70$\pm$0.17 & 16 & 15.44 & sdB \\
138.779524$^{*}$ & 18.787857 & 554312158 & 635660764741010304 & 27270$\pm$480 & 5.05$\pm$0.05 & -1.42$\pm$0.06 & 13 & 15.88 & sdB \\
140.528329$^{*}$ & 8.325203 & 223715183 & 587484528898551552 & 32160$\pm$450 & 6.01$\pm$0.12 & -1.77$\pm$0.08 & 17 & 15.59 & sdB \\
140.978208 & 36.393264 & 213108067 & 810691780249660800 & 34210$\pm$260 & 5.62$\pm$0.05 & -1.70$\pm$0.13 & 19 & 15.66 & sdOB \\
141.167117$^{*}$ & 30.836992$^{\dagger}$ & 189207170 & 700362419831670016 & 40480$\pm$130 & 5.46$\pm$0.24 & 0.82$\pm$0.04 & 75 & 14.57 & He-sdOB \\
141.196371$^{*}$ & 23.321494 & 275512146 & 639171814606295424 & 26010$\pm$210 & 5.49$\pm$0.02 & -2.29$\pm$0.10 & 11 & 15.78 & sdB \\
141.660741$^{*}$ & 32.753077 & 434508092 & 701197911230156672 & 31700$\pm$150 & 5.82$\pm$0.04 & -2.89$\pm$0.24 & 34 & 14.88 & sdB \\
142.694923$^{*}$ & 48.273269 & 208410025 & 825628714431805056 & 45830$\pm$450 & 5.78$\pm$0.06 & 0.75$\pm$0.11 & 56 & 10.64 & He-sdOB \\
143.191496$^{*}$ & 8.271814 & 219713039 & 588040469465211904 & 31710$\pm$420 & 5.73$\pm$0.04 & -1.84$\pm$0.08 & 10 & 16.47 & sdB \\
143.800665$^{*}$ & 31.166788 & 22603091$^{\ddagger}$ & 697707958245002880 & 33660$\pm$280 & 5.85$\pm$0.05 & -1.73$\pm$0.04 & 29 & 15.63 & sdOB \\
143.922229$^{*}$ & 16.353044 & 32106171 & 619153556155078272 & 30320$\pm$40 & 5.47$\pm$0.02 & -3.00> & 26 & 14.72 & sdB \\
144.249525$^{*}$ & 38.122706 & 342411212 & 800799439774706560 & 26820$\pm$880 & 5.60$\pm$0.08 & -2.40$\pm$0.11 & 11 & 15.43 & sdB \\
144.3178$^{*}$ & 18.419783 & 142410002$^{\ddagger}$ & 633011250955383168 & 35260$\pm$200 & 5.60$\pm$0.02 & -2.76> & 59 & 13.10 & sdB \\
144.584797$^{*}$ & 55.097246 & 124705056$^{\ddagger}$ & 1021648582279419648 & 32480$\pm$390 & 7.06$\pm$0.07 & -1.85$\pm$0.71 & 50 & 12.03 & He-sdOB \\
145.729342$^{*}$ & 6.593786 & 341903193 & 3853346290335592320 & 26740$\pm$380 & 5.51$\pm$0.03 & -2.82$\pm$0.10 & 38 & 13.68 & sdB \\
145.81592$^{*}$ & 16.932532 & 526902074 & 620566974057395840 & 28560$\pm$540 & 5.54$\pm$0.08 & -2.23$\pm$0.07 & 18 & 16.17 & sdB \\
145.977337$^{*}$ & 27.783209 & 447314049 & 648029239761019776 & 30570$\pm$260 & 5.51$\pm$0.01 & -3.00> & 89 & 13.21 & sdB \\
146.188903$^{*}$ & 46.592324 & 301015204 & 821716652060662144 & 33620$\pm$400 & 5.21$\pm$0.06 & -2.60$\pm$0.16 & 15 & 16.80 & sdB \\
146.872334$^{*}$ & 27.274173 & 388912003$^{\ddagger}$ & 647193232966817280 & 29550$\pm$320 & 5.57$\pm$0.06 & -2.88> & 14 & 16.63 & sdB \\
147.23733$^{*}$ & 33.697501 & 188204008 & 793720783914955392 & 51270$\pm$470 & 5.94$\pm$0.09 & -0.00$\pm$0.05 & 27 & 17.69 & He-sdOB \\
147.74187$^{*}$ & 18.438487 & 142508056$^{\ddagger}$ & 627130616093753600 & 34560$\pm$250 & 5.89$\pm$0.08 & -1.78$\pm$0.06 & 12 & 16.12 & sdOB \\
148.383114$^{*}$ & 11.750013 & 406114080 & 613219762482485632 & 34780$\pm$510 & 5.83$\pm$0.11 & -1.49$\pm$0.10 & 32 & 15.10 & sdOB \\
148.831108 & 51.616397 & 293910209 & 828120310859359104 & 49140$\pm$860 & 5.61$\pm$0.13 & 0.78$\pm$0.21 & 74 & 12.61 & He-sdOB \\
149.202242$^{*}$ & 14.378333 & 210808231 & 615674873163301248 & 45520$\pm$180 & 5.24$\pm$0.07 & 0.84$\pm$0.09 & 69 & 14.31 & He-sdOB \\
149.444189$^{*}$ & 4.913523 & 445716215 & 3849270435091240192 & 26390$\pm$140 & 5.49$\pm$0.01 & -2.45$\pm$0.03 & 31 & 15.32 & sdB \\
149.966705$^{*}$ & 3.50916 & 445703068$^{\ddagger}$ & 3848520258923494656 & 33770$\pm$220 & 5.30$\pm$0.03 & -2.60$\pm$0.14 & 44 & 15.41 & sdOB \\
149.98337$^{*}$ & 35.666425 & 90910097 & 795785941630955392 & 46650$\pm$1320 & 6.11$\pm$0.15 & -0.36$\pm$0.14 & 25 & 15.33 & He-sdOB \\
153.42551$^{*}$ & 26.10556 & 134015249$^{\ddagger}$ & 739156243780100352 & 52050$\pm$2820 & 5.37$\pm$0.08 & -1.73$\pm$0.28 & 10 & 16.67 & sdO \\
153.7891$^{*}$ & 0.550186 & 315411004 & 3831905500061334912 & 32980$\pm$360 & 5.84$\pm$0.07 & -2.06$\pm$0.08 & 21 & 14.86 & sdOB \\
155.124196$^{*}$ & 42.839426 & 34604022$^{\ddagger}$ & 805690720332579456 & 39670$\pm$690 & 5.25$\pm$0.07 & -2.54$\pm$0.19 & 32 & 15.28 & sdO \\
158.819075$^{*}$ & 40.354013 & 142615080$^{\ddagger}$ & 780253140863277184 & 31660$\pm$110 & 5.92$\pm$0.04 & -2.24$\pm$0.02 & 93 & 11.46 & sdB \\
159.16221$^{*}$ & 19.867303 & 106016182 & 3987249901607922944 & 33980$\pm$450 & 5.80$\pm$0.04 & -1.78$\pm$0.06 & 20 & 15.59 & sdOB \\
160.346854$^{*}$ & 50.738881$^{\dagger}$ & 18316005$^{\ddagger}$ & 836644480912524288 & 50230$\pm$1360 & 5.70$\pm$0.09 & 0.22$\pm$0.07 & 20 & 14.95 & He-sdO \\
160.376837$^{*}$ & 18.703028$^{\dagger}$ & 296504043 & 3986138020473930368 & 34890$\pm$520 & 5.98$\pm$0.03 & -1.68$\pm$0.03 & 61 & 13.00 & sdOB \\
163.185161$^{*}$ & 24.990381 & 403016118 & 729598330934052480 & 33540$\pm$490 & 5.81$\pm$0.14 & -1.90$\pm$0.06 & 18 & 16.73 & sdOB \\
163.553054$^{*}$ & 29.286844 & 392414214 & 731743061507469568 & 35890$\pm$610 & 5.96$\pm$0.08 & -0.63$\pm$0.09 & 18 & 17.08 & He-sdOB \\
163.5772$^{*}$ & 49.833258 & 148604206$^{\ddagger}$ & 832980633290384256 & 34250$\pm$1180 & 5.21$\pm$0.05 & -1.53$\pm$0.04 & 56 & 13.34 & sdOB \\
164.49736$^{*}$ & 47.052083 & 504010077 & 831501240400109824 & 39310$\pm$1160 & 5.83$\pm$0.18 & 1.86$\pm$0.07 & 11 & 17.23 & He-sdB \\
165.233143$^{*}$ & 10.928443$^{\dagger}$ & 290014160 & 3868418219635118080 & 32320$\pm$90 & 5.93$\pm$0.05 & -1.92$\pm$0.05 & 37 & 14.21 & sdOB \\
165.75989$^{*}$ & -1.060761 & 404011209 & 3803624106785054848 & 29360$\pm$390 & 5.48$\pm$0.07 & -2.52$\pm$0.10 & 16 & 16.46 & sdB \\
166.34654$^{*}$ & 49.583249 & 503916138 & 832788184395879424 & 78590$\pm$2370 & 5.36$\pm$0.05 & -0.97$\pm$0.14 & 51 & 14.26 & sdO \\
166.7103$^{*}$ & 29.592403 & 392409050 & 732692631532422528 & 35630$\pm$400 & 5.86$\pm$0.05 & -1.48$\pm$0.04 & 44 & 15.73 & sdOB \\
166.8567 & 47.353497 & 504003005 & 783541848870827264 & 25510$\pm$890 & 5.63$\pm$0.11 & -3.00> & 19 & 16.75 & sdB \\
167.335037$^{*}$ & 26.790214 & 190610055 & 3998182792399323904 & 40170$\pm$1230 & 5.60$\pm$0.12 & -2.84$\pm$0.26 & 13 & 15.49 & sdO \\
\hline\noalign{\smallskip} 
  \end{tabularx}
\end{table*}

\setcounter{table}{0}
\begin{table*}
\tiny

 \begin{minipage}{160mm}
  \caption{Continued.}
  \end{minipage}\\
  \centering
    \begin{tabularx}{16.0cm}{lllccccccccccccccX}
\hline\noalign{\smallskip}
RA\tablenotemark{a}  & DEC\tablenotemark{b}   &obsid\tablenotemark{c} & source\_id 
& $T_\mathrm{eff}$ &  $\mathrm{log}\ g$ & $\mathrm{log}(n\mathrm{He}/n\mathrm{H})$\tablenotemark{d}
&SNRU & $G$ & spclass   \\
 LAMOST &  LAMOST& LAMOST&Gaia  &(K)&($\mathrm{cm\ s^{-2}}$)& &&Gaia(mag) & \\
\hline\noalign{\smallskip}
169.64492$^{*}$ & 19.829744 & 297308031 & 3978207002584425216 & 28090$\pm$230 & 5.42$\pm$0.01 & -2.60$\pm$0.06 & 45 & 13.17 & sdB \\
169.770304$^{*}$ & 29.864886 & 27611095$^{\ddagger}$ & 4023163971460301952 & 31520$\pm$340 & 6.19$\pm$0.07 & -2.33$\pm$0.17 & 19 & 14.30 & sdB \\
170.234298$^{*}$ & 9.611609 & 419405245 & 3914988790544222976 & 47810$\pm$6960 & 5.64$\pm$0.11 & -2.54$\pm$0.23 & 12 & 16.45 & sdO \\
170.57449$^{*}$ & 37.448018 & 439213135 & 760930907631979904 & 39140$\pm$280 & 5.45$\pm$0.06 & -2.76$\pm$0.25 & 20 & 15.93 & sdO \\
171.654513$^{*}$ & 11.999694 & 419412143$^{\ddagger}$ & 3917380026471466112 & 29340$\pm$290 & 5.44$\pm$0.06 & -3.11$\pm$0.18 & 16 & 15.71 & sdB \\
172.122092$^{*}$ & 29.251333 & 27613119$^{\ddagger}$ & 4022181695259436416 & 46740$\pm$2870 & 6.20$\pm$0.22 & -1.91$\pm$0.13 & 10 & 15.06 & sdO \\
172.798333$^{*}$ & 19.611231 & 393415146 & 3977575573672822400 & 30440$\pm$230 & 5.74$\pm$0.05 & -3.00> & 28 & 15.57 & sdB \\
173.173392$^{*}$ & -6.614569 & 407007245 & 3593223417120115712 & 49200$\pm$1400 & 5.73$\pm$0.07 & -2.92$\pm$0.26 & 17 & 16.24 & sdO \\
173.41895$^{*}$ & 56.106739 & 42512139$^{\ddagger}$ & 844818765749387392 & 32380$\pm$720 & 5.11$\pm$0.08 & -2.81$\pm$0.15 & 11 & 15.21 & sdB \\
174.359894$^{*}$ & 14.170597 & 571607128 & 3918117008499710848 & 35700$\pm$820 & 5.83$\pm$0.09 & -0.38$\pm$0.04 & 33 & 13.25 & He-sdOB \\
174.47542$^{*}$ & 58.256857 & 449009063 & 846070525377462528 & 33830$\pm$1110 & 5.88$\pm$0.05 & -2.27$\pm$0.10 & 22 & 16.18 & sdOB \\
174.925071$^{*}$ & 46.730407 & 140606089$^{\ddagger}$ & 786220877662239488 & 30760$\pm$170 & 5.55$\pm$0.05 & -3.00> & 25 & 15.57 & sdB \\
176.238472$^{*}$ & -3.948095 & 499808226 & 3600365496761173504 & 28010$\pm$340 & 5.31$\pm$0.06 & -2.88$\pm$0.13 & 21 & 15.88 & sdB \\
179.001258$^{*}$ & 34.123619 & 399012110 & 4028265945931548544 & 38180$\pm$310 & 5.42$\pm$0.05 & -2.54$\pm$0.09 & 58 & 14.83 & sdOB \\
180.921546$^{*}$ & 25.519844 & 19002085 & 4006102059258774272 & 35060$\pm$630 & 5.63$\pm$0.13 & -1.53$\pm$0.10 & 13 & 15.13 & sdOB \\
182.250307 & 47.915233 & 448510114 & 1546073572415417856 & 25900$\pm$450 & 5.29$\pm$0.06 & -1.97$\pm$0.09 & 14 & 16.61 & sdB \\
182.319787$^{*}$ & 16.198908 & 343510162 & 3922570889586104064 & 29840$\pm$60 & 5.67$\pm$0.06 & -2.41$\pm$0.03 & 84 & 13.65 & sdB \\
182.400326$^{*}$ & -3.552102 & 203010225 & 3598151393876907904 & 36140$\pm$850 & 5.88$\pm$0.02 & -1.56$\pm$0.04 & 48 & 13.34 & sdOB \\
182.953929$^{*}$ & 16.562911 & 404103003 & 3922610192830905984 & 68520$\pm$25560 & 5.64$\pm$0.24 & -1.89> & 17 & 16.61 & sdO \\
189.77099$^{*}$ & 47.630772 & 319211089 & 1543006141129419648 & 29650$\pm$100 & 5.75$\pm$0.20 & -2.33$\pm$0.12 & 11 & 15.56 & sdB \\
190.026263$^{*}$ & 11.538483 & 220712168 & 3928326210186248832 & 34620$\pm$340 & 5.28$\pm$0.08 & -2.43$\pm$0.22 & 11 & 16.00 & sdB \\
190.630828$^{*}$ & 11.733614 & 437602078 & 3928420360164984320 & 47940$\pm$690 & 5.56$\pm$0.23 & 0.75$\pm$0.12 & 50 & 15.71 & He-sdOB \\
190.70374$^{*}$ & 13.609056 & 437603053 & 3929230528435328640 & 34320$\pm$510 & 5.05$\pm$0.08 & -3.15$\pm$0.79 & 13 & 17.59 & sdB \\
192.0795$^{*}$ & 3.834201 & 437115236 & 3703904449460695040 & 74290$\pm$18350 & 5.85$\pm$0.09 & -1.32$\pm$0.08 & 20 & 16.83 & sdO \\
193.034742$^{*}$ & 11.850803 & 223907015 & 3927916302803458304 & 31600$\pm$280 & 5.63$\pm$0.05 & -2.41$\pm$0.10 & 20 & 15.25 & sdB \\
193.12335$^{*}$ & -3.024892 & 105710231$^{\ddagger}$ & 3682304199935092096 & 29880$\pm$160 & 5.72$\pm$0.03 & -3.00> & 15 & 15.65 & sdB \\
193.32712$^{*}$ & 30.108144 & 144516050$^{\ddagger}$ & 1465043619891174016 & 33350$\pm$140 & 5.68$\pm$0.06 & -2.05$\pm$0.08 & 16 & 15.84 & sdB \\
193.534896$^{*}$ & 1.723411 & 212208212 & 3690648771635312512 & 50770$\pm$910 & 6.03$\pm$0.09 & 0.01$\pm$0.06 & 14 & 15.53 & He-sdOB \\
194.114387$^{*}$ & 27.708522 & 301803019$^{\ddagger}$ & 1463697233542956672 & 26330$\pm$160 & 5.75$\pm$0.06 & -3.00> & 28 & 15.82 & sdB \\
195.246685$^{*}$ & 0.953267 & 327515053 & 3689564206493682816 & 39660$\pm$410 & 5.70$\pm$0.04 & -0.49$\pm$0.06 & 18 & 16.47 & He-sdOB \\
195.944218$^{*}$ & 26.775245 & 301808222$^{\ddagger}$ & 1460356814139885568 & 45940$\pm$230 & 6.93$\pm$0.01 & -0.73$\pm$0.12 & 42 & 15.60 & He-sdOB \\
196.202858$^{*}$ & 28.124981 & 34503131$^{\ddagger}$ & 1461140216174118016 & 34360$\pm$1240 & 5.97$\pm$0.08 & -3.41$\pm$0.57 & 13 & 15.46 & sdB \\
196.433237$^{*}$ & 11.978208 & 214502083 & 3737057611255721472 & 31840$\pm$290 & 5.63$\pm$0.05 & -2.32$\pm$0.15 & 18 & 16.04 & sdB \\
196.564867$^{*}$ & 48.838822 & 149710037$^{\ddagger}$ & 1554816510916988544 & 32570$\pm$110 & 5.69$\pm$0.02 & -1.76$\pm$0.02 & 52 & 13.73 & sdOB \\
198.206285$^{*}$ & 54.545775 & 403709179 & 1564081030251770752 & 28560$\pm$510 & 5.34$\pm$0.05 & -2.61$\pm$0.17 & 14 & 15.84 & sdB \\
199.822539$^{*}$ & 20.95443 & 401811071 & 3940518836361346304 & 51410$\pm$980 & 5.50$\pm$0.05 & 0.82$\pm$0.12 & 49 & 15.15 & He-sdOB \\
200.184944$^{*}$ & 5.983701 & 36109220$^{\ddagger}$ & 3717383603022961792 & 43660$\pm$610 & 5.70$\pm$0.10 & 0.70$\pm$0.16 & 12 & 14.74 & He-sdOB \\
200.848058$^{*}$ & 26.275522 & 582314065 & 1449056067988016128 & 33770$\pm$480 & 5.89$\pm$0.07 & -1.71$\pm$0.06 & 26 & 16.19 & sdOB \\
200.896934$^{*}$ & 36.133203 & 139410221 & 1473687671071803520 & 36900$\pm$480 & 5.47$\pm$0.04 & -0.20$\pm$0.72 & 109 & 11.64 & He-sdOB \\
200.988725$^{*}$ & 26.250672 & 219310248 & 1449053353568686336 & 27330$\pm$780 & 5.21$\pm$0.09 & -2.21$\pm$0.14 & 11 & 17.31 & sdB \\
201.134907 & 32.072558 & 202506130 & 1469357759922416256 & 22580$\pm$590 & 5.06$\pm$0.06 & -2.17$\pm$0.11 & 15 & 16.64 & sdB \\
205.38117$^{*}$ & 4.912973 & 142010081$^{\ddagger}$ & 3714909426982422272 & 59430$\pm$4100 & 6.38$\pm$0.11 & -1.59$\pm$0.19 & 11 & 16.23 & sdO \\
206.636362$^{*}$ & 28.289811 & 448603239 & 1451833434359977216 & 28040$\pm$360 & 5.33$\pm$0.02 & -2.93$\pm$0.07 & 68 & 14.87 & sdB \\
207.566067$^{*}$ & 60.410686 & 36709003$^{\ddagger}$ & 1661062937982961664 & 54020$\pm$1310 & 5.62$\pm$0.15 & -0.43$\pm$0.07 & 16 & 16.32 & He-sdO \\
207.971317$^{*}$ & -1.496278 & 41804169$^{\ddagger}$ & 3658738470295900800 & 30420$\pm$720 & 5.70$\pm$0.06 & -3.39$\pm$0.23 & 10 & 15.66 & sdB \\
208.89461$^{*}$ & 16.004774 & 341310209 & 1243253467230351360 & 37380$\pm$390 & 5.69$\pm$0.05 & -1.38$\pm$0.06 & 33 & 15.92 & sdOB \\
208.913404$^{*}$ & 14.910708 & 341302147 & 1231024630186706944 & 29210$\pm$190 & 5.48$\pm$0.04 & -2.39$\pm$0.06 & 27 & 16.00 & sdB \\
209.602612$^{*}$ & 6.860208$^{\dagger}$ & 339610163$^{\ddagger}$ & 3720655474749294592 & 28470$\pm$380 & 5.41$\pm$0.08 & -2.90$\pm$0.08 & 65 & 14.32 & sdB \\
210.821432$^{*}$ & 28.658128 & 332003169 & 1452740084776034944 & 47360$\pm$490 & 5.57$\pm$0.05 & 0.58$\pm$0.09 & 76 & 14.85 & He-sdOB \\
211.499225$^{*}$ & 31.409769 & 537808228 & 1454323415879594368 & 30540$\pm$220 & 5.73$\pm$0.04 & -2.33$\pm$0.07 & 48 & 13.48 & sdB \\
213.598658$^{*}$ & 29.684781 & 233014059 & 1260997037986811648 & 24980$\pm$300 & 5.68$\pm$0.04 & -3.00> & 11 & 16.66 & sdB \\
213.772692$^{*}$ & 27.459303 & 233002202 & 1260448626498096000 & 28630$\pm$230 & 5.47$\pm$0.03 & -2.40$\pm$0.13 & 19 & 16.72 & sdB \\
214.401671$^{*}$ & -4.574742 & 115709085$^{\ddagger}$ & 3642680992030225536 & 36370$\pm$390 & 5.79$\pm$0.07 & -1.48$\pm$0.05 & 24 & 13.71 & sdOB \\
214.587163$^{*}$ & -3.381686 & 449706090 & 3645920088861401216 & 29790$\pm$290 & 5.79$\pm$0.03 & -3.00> & 17 & 16.51 & sdB \\
214.84398$^{*}$ & 29.780611 & 233003129 & 1284970278417991680 & 34740$\pm$350 & 5.62$\pm$0.11 & -1.52$\pm$0.06 & 20 & 17.40 & sdOB \\
216.248249$^{*}$ & 3.328705 & 236906092 & 3668499109893566464 & 35200$\pm$330 & 5.82$\pm$0.32 & -1.49$\pm$0.13 & 14 & 16.48 & sdOB \\
216.49655$^{*}$ & 28.787578 & 233008016 & 1283953333241515520 & 34250$\pm$200 & 5.64$\pm$0.02 & -1.78$\pm$0.05 & 15 & 16.74 & sdB \\
217.365692$^{*}$ & 19.359703 & 316813056 & 1239364769480858496 & 50580$\pm$1460 & 5.55$\pm$0.16 & 0.29$\pm$0.06 & 39 & 14.08 & He-sdOB \\
218.77645$^{*}$ & 28.797477 & 566705209 & 1281101268798929152 & 36500$\pm$320 & 5.85$\pm$0.05 & -1.77$\pm$0.05 & 27 & 16.10 & sdOB \\
221.04239$^{*}$ & 34.353706 & 574013227 & 1292915830478091008 & 46830$\pm$310 & 5.72$\pm$0.17 & 1.02$\pm$0.12 & 38 & 16.03 & He-sdOB \\
221.784479$^{*}$ & 7.397089 & 140208168$^{\ddagger}$ & 1171708967165771392 & 49380$\pm$890 & 5.43$\pm$0.15 & 0.78$\pm$0.07 & 32 & 14.63 & He-sdOB \\
223.344967$^{*}$ & 39.496492 & 242506213 & 1296418427846820608 & 30110$\pm$470 & 5.81$\pm$0.11 & -2.68$\pm$0.28 & 15 & 16.16 & sdB \\
223.61111$^{*}$ & 47.334557 & 344015107 & 1590388529573233536 & 30350$\pm$40 & 5.74$\pm$0.02 & -2.50$\pm$0.09 & 28 & 16.11 & sdB \\
224.0603$^{*}$ & 16.961273 & 566108203 & 1188049561085487872 & 38920$\pm$950 & 5.76$\pm$0.29 & 2.07$\pm$0.13 & 22 & 17.28 & He-sdB \\
224.39335$^{*}$ & 3.79988 & 457101145 & 1155001746247388800 & 29520$\pm$630 & 5.68$\pm$0.06 & -2.58$\pm$0.14 & 14 & 16.76 & sdB \\
226.152825$^{*}$ & 12.803728 & 442407239 & 1181028908759001472 & 46180$\pm$460 & 5.48$\pm$0.09 & 0.52$\pm$0.09 & 40 & 15.45 & He-sdOB \\
229.805583$^{*}$ & 26.289058 & 462202202 & 1270423258548105728 & 30470$\pm$190 & 5.46$\pm$0.09 & -3.00> & 15 & 15.86 & sdB \\
231.759504$^{*}$ & 42.068681 & 573415155 & 1390836579404946560 & 30860$\pm$150 & 5.73$\pm$0.05 & -2.90$\pm$0.15 & 24 & 16.31 & sdB \\
231.77099$^{*}$ & 11.14553 & 565710140 & 1169270800131487360 & 36990$\pm$290 & 5.61$\pm$0.04 & -0.45$\pm$0.04 & 26 & 17.24 & He-sdOB \\
232.084522$^{*}$ & 10.50891 & 567310064 & 1165848673269627520 & 35120$\pm$320 & 5.75$\pm$0.06 & -1.36$\pm$0.04 & 21 & 15.95 & sdOB \\
232.217458$^{*}$ & 9.528992 & 567302092 & 1165536789924111616 & 32720$\pm$170 & 5.88$\pm$0.04 & -1.87$\pm$0.06 & 17 & 16.45 & sdOB \\
232.746779$^{*}$ & 2.306311 & 557703122 & 4421020604704127360 & 88700$\pm$29720 & 5.21$\pm$0.39 & 1.79$\pm$0.20 & 12 & 16.43 & He-sdO \\
233.071679$^{*}$ & 45.772572 & 582416115 & 1394898866848350848 & 34680$\pm$680 & 5.35$\pm$0.22 & -1.88$\pm$0.14 & 10 & 16.54 & sdOB \\
233.374829$^{*}$ & 52.113533 & 53103086$^{\ddagger}$ & 1595357428778377344 & 31890$\pm$270 & 6.04$\pm$0.06 & -2.38$\pm$0.09 & 12 & 13.99 & sdB \\
233.796725 & 23.788525 & 328505074 & 1220934996098999552 & 27150$\pm$260 & 5.37$\pm$0.03 & -2.87$\pm$0.10 & 18 & 15.58 & sdB \\
235.556658$^{*}$ & 4.143363 & 449204160 & 4424645557102354432 & 68620$\pm$4760 & 5.17$\pm$0.23 & 0.86$\pm$0.58 & 21 & 15.90 & He-sdO \\
236.678871 & 0.757753 & 457502165 & 4422600328035803264 & 52670$\pm$1420 & 5.43$\pm$0.12 & -2.80$\pm$0.30 & 19 & 15.11 & sdO \\
237.67498$^{*}$ & 29.756532 & 339709096 & 1320813464131916416 & 30770$\pm$60 & 5.53$\pm$0.05 & -2.69$\pm$0.08 & 34 & 16.41 & sdB \\
239.140379$^{*}$ & 22.046203 & 346508084 & 1206214356465503616 & 78490$\pm$9580 & 5.20$\pm$0.06 & -0.16$\pm$0.03 & 36 & 13.81 & He-sdO \\
\hline\noalign{\smallskip} 
  \end{tabularx}
\end{table*}

\setcounter{table}{0}
\begin{table*}
\tiny

 \begin{minipage}{160mm}
  \caption{Continued.}
  \end{minipage}\\
  \centering
    \begin{tabularx}{16.0cm}{lllcccccccccccccX}
\hline\noalign{\smallskip}
RA\tablenotemark{a}  & DEC\tablenotemark{b}   &obsid\tablenotemark{c} & source\_id 
& $T_\mathrm{eff}$ &  $\mathrm{log}\ g$ & $\mathrm{log}(n\mathrm{He}/n\mathrm{H})$\tablenotemark{d}
&SNRU & $G$ & spclass \\
 LAMOST &  LAMOST& LAMOST&Gaia  &(K)&($\mathrm{cm\ s^{-2}}$)& &&Gaia(mag) & \\
\hline\noalign{\smallskip}
239.178931$^{*}$ & 50.26042 & 437204128 & 1403149533383468928 & 63150$\pm$2470 & 6.54$\pm$0.10 & -0.41$\pm$0.20 & 25 & 16.13 & He-sdO \\
239.374454$^{*}$ & 48.839656 & 437201058 & 1399993557054575232 & 43920$\pm$610 & 6.01$\pm$0.13 & 0.63$\pm$0.09 & 13 & 16.52 & He-sdOB \\
239.69352$^{*}$ & 19.371987 & 558416059 & 1203338412004998016 & 61690$\pm$9680 & 5.67$\pm$0.11 & -1.39$\pm$0.07 & 28 & 16.81 & sdO \\
240.88683$^{*}$ & 51.80773 & 458002156 & 1403885518979474688 & 35500$\pm$640 & 5.92$\pm$0.09 & -0.61$\pm$0.23 & 12 & 17.75 & He-sdOB \\
242.385059$^{*}$ & 22.630508 & 461109174 & 1206447693448896128 & 44740$\pm$530 & 5.64$\pm$0.11 & 1.89$\pm$0.10 & 20 & 16.34 & He-sdOB \\
242.88709$^{*}$ & 29.01083 & 342203079 & 1316998296222144768 & 30080$\pm$750 & 5.41$\pm$0.15 & -2.66$\pm$0.17 & 12 & 16.83 & sdB \\
242.95559$^{*}$ & 19.397395 & 558412106 & 1201659389029469184 & 33090$\pm$170 & 5.81$\pm$0.05 & -1.93$\pm$0.03 & 17 & 15.89 & sdOB \\
243.365233$^{*}$ & 8.898758$^{\dagger}$ & 233106152 & 4452964853466754048 & 31400$\pm$140 & 5.92$\pm$0.04 & -2.19$\pm$0.06 & 90 & 14.63 & sdB \\
243.572692$^{*}$ & 3.997775 & 568307134 & 4437211291383115392 & 46690$\pm$850 & 5.60$\pm$0.05 & 0.65$\pm$0.06 & 23 & 15.58 & He-sdOB \\
243.696$^{*}$ & 42.459869 & 241910003 & 1381954385173920640 & 35200$\pm$1410 & 5.83$\pm$0.08 & -1.53$\pm$0.05 & 30 & 14.38 & sdOB \\
244.13037$^{*}$ & -0.648126 & 451409063 & 4407283954965874816 & 36490$\pm$690 & 5.53$\pm$0.08 & -1.68$\pm$0.11 & 11 & 16.82 & sdOB \\
244.84043$^{*}$ & 29.918343 & 342209165 & 1318236621192489216 & 33060$\pm$290 & 5.93$\pm$0.13 & -2.12$\pm$0.14 & 10 & 16.87 & sdB \\
245.874442$^{*}$ & 24.761389 & 238215199 & 1302449042967968640 & 53080$\pm$740 & 5.68$\pm$0.08 & 0.21$\pm$0.06 & 37 & 15.70 & He-sdO \\
246.60941$^{*}$ & 33.554169 & 344109041 & 1325657289591907456 & 44640$\pm$590 & 5.34$\pm$0.09 & 1.20$\pm$0.45 & 11 & 18.25 & He-sdOB \\
246.725946$^{*}$ & 8.426436 & 236512210 & 4440350981196959104 & 41520$\pm$40 & 5.64$\pm$0.04 & 1.10$\pm$0.05 & 46 & 14.91 & He-sdOB \\
248.05109$^{*}$ & 17.888424 & 146615140$^{\ddagger}$ & 4467169787782385664 & 36470$\pm$1150 & 5.62$\pm$0.24 & -2.75$\pm$0.57 & 11 & 15.99 & sdB \\
248.647179$^{*}$ & 26.350464 & 566608204 & 1304379201270414592 & 51420$\pm$1060 & 5.70$\pm$0.11 & -2.31$\pm$0.11 & 20 & 16.97 & sdO \\
251.538521$^{*}$ & 40.290433 & 152701066$^{\ddagger}$ & 1355775219474985856 & 29810$\pm$130 & 5.69$\pm$0.01 & -1.93$\pm$0.05 & 40 & 14.10 & sdB \\
253.7016$^{*}$ & 15.882959 & 546709115 & 4557895027682278784 & 43060$\pm$360 & 5.59$\pm$0.10 & 0.75$\pm$0.08 & 38 & 15.53 & He-sdOB \\
255.65702$^{*}$ & 24.589606 & 143802076$^{\ddagger}$ & 4572469019942466304 & 27590$\pm$690 & 5.46$\pm$0.06 & -2.38$\pm$0.12 & 10 & 15.93 & sdB \\
255.98617$^{*}$ & 34.251398 & 565812178 & 1337803736279119616 & 28360$\pm$180 & 5.34$\pm$0.03 & -2.54$\pm$0.12 & 14 & 16.72 & sdB \\
256.02866$^{*}$ & 28.90447 & 346705170 & 1308731137012654464 & 32720$\pm$320 & 5.92$\pm$0.07 & -1.69$\pm$0.07 & 15 & 17.04 & sdOB \\
256.04676$^{*}$ & 29.596299 & 346704165 & 1308836793208539264 & 30660$\pm$350 & 5.53$\pm$0.05 & -2.63$\pm$0.14 & 23 & 17.16 & sdB \\
256.313389$^{*}$ & 35.471812 & 576815103 & 1338485777087459072 & 43090$\pm$240 & 5.79$\pm$0.03 & 0.91$\pm$0.08 & 42 & 15.58 & He-sdOB \\
256.39427$^{*}$ & 24.890831 & 143802233$^{\ddagger}$ & 4572125628717747200 & 34410$\pm$620 & 5.80$\pm$0.07 & -1.58$\pm$0.07 & 10 & 16.86 & sdB \\
258.269431 & 24.838763 & 564407046 & 4571972212486265216 & 27400$\pm$390 & 5.44$\pm$0.02 & -2.84$\pm$0.34 & 11 & 17.07 & sdB \\
258.790469 & 24.739852 & 564407018 & 4573406491044824448 & 34770$\pm$750 & 5.19$\pm$0.08 & -2.82$\pm$0.26 & 10 & 16.93 & sdO \\
259.37476$^{*}$ & 32.12192 & 573808096 & 1333535634003198848 & 32540$\pm$410 & 5.76$\pm$0.07 & -2.28$\pm$0.16 & 19 & 17.23 & sdB \\
261.050108$^{*}$ & 28.591214 & 227215215 & 4598511330803479936 & 36350$\pm$210 & 5.89$\pm$0.03 & -1.67$\pm$0.03 & 77 & 13.34 & sdOB \\
262.794308 & 53.086861 & 462011209 & 1417526777492843136 & 27550$\pm$530 & 5.17$\pm$0.06 & -2.64$\pm$0.21 & 12 & 17.02 & sdB \\
265.676651 & 44.503737 & 568416221 & 1350113181268869632 & 37500$\pm$480 & 5.56$\pm$0.15 & -2.41$\pm$0.23 & 11 & 16.02 & sdOB \\
272.615979$^{*}$ & 41.90375 & 239111121 & 2113506200378734848 & 27760$\pm$190 & 5.37$\pm$0.03 & -2.86$\pm$0.06 & 27 & 16.15 & sdB \\
279.219563 & 57.462023 & 346803047 & 2154234138295606272 & 28290$\pm$300 & 5.71$\pm$0.07 & -3.00> & 15 & 16.47 & sdB \\
281.163262 & 43.374406 & 457204145 & 2116887915892603520 & 28740$\pm$1030 & 5.83$\pm$0.17 & -1.89$\pm$0.17 & 37 & 16.30 & sdB \\
294.50684$^{*}$ & 46.82906 & 354601052 & 2128531679610988672 & 40680$\pm$20 & 5.17$\pm$0.15 & 1.59$\pm$0.07 & 23 & 15.98 & He-sdOB \\
314.217996$^{*}$ & 3.596911$^{\dagger}$ & 370216014 & 1731435633531386752 & 40220$\pm$560 & 5.27$\pm$0.04 & -3.25$\pm$0.27 & 34 & 14.46 & sdO \\
315.813279 & 4.684183 & 370211146 & 1733071908696066176 & 29660$\pm$320 & 5.69$\pm$0.06 & -2.85$\pm$0.28 & 17 & 16.58 & sdB \\
317.098329$^{*}$ & 1.032085 & 370208205 & 2690647825256267264 & 25480$\pm$500 & 5.33$\pm$0.04 & -2.50$\pm$0.06 & 26 & 15.99 & sdB \\
318.667075 & 20.911639 & 372014148 & 1789867186124831616 & 29020$\pm$210 & 5.32$\pm$0.03 & -3.00> & 10 & 16.38 & sdB \\
319.635091 & 23.571516 & 258303199 & 1792255325381143040 & 43100$\pm$240 & 5.43$\pm$0.04 & 0.77$\pm$0.12 & 67 & 14.67 & He-sdOB \\
319.858333$^{*}$ & 20.365419 & 469903131 & 1790470508770502144 & 43130$\pm$340 & 5.58$\pm$0.17 & 0.93$\pm$0.10 & 45 & 15.99 & He-sdOB \\
320.456296$^{*}$ & 20.371406 & 372015115 & 1790305753825533312 & 39600$\pm$320 & 5.33$\pm$0.06 & -2.94$\pm$0.17 & 28 & 15.37 & sdO \\
321.270667$^{*}$ & -0.035192 & 254804129 & 2687886088269786112 & 36740$\pm$580 & 5.73$\pm$0.08 & -1.43$\pm$0.07 & 22 & 17.02 & sdOB \\
321.356173$^{*}$ & 8.319926 & 372503172 & 1740252956575806848 & 33230$\pm$260 & 5.53$\pm$0.18 & -1.33$\pm$0.28 & 11 & 16.59 & sdOB \\
321.9556$^{*}$ & 0.490044 & 254804227 & 2687977931850851584 & 58450$\pm$10450 & 6.36$\pm$0.10 & -1.80$\pm$0.19 & 13 & 17.98 & sdO \\
323.510983$^{*}$ & 16.573053 & 158213246 & 1772138591918813056 & 37680$\pm$570 & 5.85$\pm$0.03 & -3.04$\pm$0.17 & 34 & 14.44 & sdO \\
323.858475$^{*}$ & -6.962075 & 56403085$^{\ddagger}$ & 2667642399870553984 & 25510$\pm$480 & 5.21$\pm$0.02 & -2.63$\pm$0.11 & 13 & 13.50 & sdB \\
325.608836 & 9.218803 & 369411090 & 1741352296404629888 & 43970$\pm$570 & 5.91$\pm$0.12 & 0.46$\pm$0.09 & 15 & 17.57 & He-sdOB \\
326.420157 & -1.468957 & 266105200 & 2674696247998989184 & 47260$\pm$360 & 5.72$\pm$0.11 & 0.37$\pm$0.05 & 36 & 15.26 & He-sdOB \\
328.835563 & 11.297908 & 371315006 & 2726863230013742336 & 35470$\pm$130 & 5.80$\pm$0.04 & -1.44$\pm$0.02 & 42 & 15.35 & sdOB \\
329.345461$^{*}$ & 11.985671 & 371311070 & 2728427319664072704 & 36080$\pm$300 & 5.93$\pm$0.06 & -2.09$\pm$0.15 & 18 & 16.47 & sdOB \\
329.4386 & -2.025295 & 266107138 & 2679677791586720896 & 29290$\pm$100 & 5.51$\pm$0.04 & -2.79$\pm$0.12 & 13 & 16.52 & sdB \\
329.725771$^{*}$ & 21.409014 & 169010006 & 1781080159778185472 & 29350$\pm$310 & 5.30$\pm$0.10 & -2.78> & 11 & 16.26 & sdB \\
329.98546 & 14.30136 & 378803208 & 1768547964899107584 & 34370$\pm$700 & 5.92$\pm$0.04 & -0.73$\pm$0.04 & 42 & 16.20 & He-sdOB \\
331.015786$^{*}$ & 14.777294 & 378809160 & 1768493259900525312 & 42900$\pm$820 & 5.65$\pm$0.08 & 1.08$\pm$0.05 & 44 & 15.79 & He-sdOB \\
331.115839 & 4.936688 & 353607003 & 2696218874579711744 & 27460$\pm$730 & 5.79$\pm$0.07 & -2.73$\pm$0.17 & 12 & 16.84 & sdB \\
331.48165$^{*}$ & 5.959966 & 353606228 & 2720511213901083264 & 73690$\pm$4430 & 6.15$\pm$0.11 & -1.80$\pm$0.28 & 11 & 18.12 & sdO \\
331.818739$^{*}$ & 3.705491 & 61309195$^{\ddagger}$ & 2684009283284602880 & 32180$\pm$130 & 5.97$\pm$0.05 & -1.93$\pm$0.12 & 12 & 14.21 & sdB \\
332.002469$^{*}$ & 2.562084 & 381708131$^{\ddagger}$ & 2683585696429935488 & 27610$\pm$300 & 5.62$\pm$0.01 & -3.00> & 53 & 14.12 & sdB \\
332.08127$^{*}$ & 6.048748 & 353606189 & 2720410333709020160 & 46210$\pm$250 & 5.35$\pm$0.07 & 1.09$\pm$0.08 & 13 & 16.73 & He-sdOB \\
332.776303 & 51.642346 & 503412047 & 2004114070560436736 & 33090$\pm$780 & 5.40$\pm$0.08 & -2.69$\pm$0.24 & 13 & 15.20 & sdB \\
333.826383 & 16.617652 & 505803150 & 1775141422597602176 & 50800$\pm$1280 & 5.53$\pm$0.05 & -1.76$\pm$0.07 & 45 & 16.87 & sdO \\
333.962114 & 1.60178 & 382503103 & 2679494173144971136 & 29590$\pm$420 & 5.33$\pm$0.07 & -2.53$\pm$0.16 & 14 & 16.69 & sdB \\
334.029951 & 15.060714 & 505805092 & 2735467282163538944 & 34620$\pm$600 & 5.89$\pm$0.06 & -1.46$\pm$0.06 & 17 & 16.98 & sdOB \\
334.07485$^{*}$ & 13.85585 & 470416164 & 2734200064947455232 & 23150$\pm$260 & 5.05$\pm$0.05 & -2.19$\pm$0.03 & 19 & 15.83 & sdB \\
334.529784 & 4.819463 & 368707127 & 2707980797138700672 & 51910$\pm$1320 & 5.41$\pm$0.07 & 0.25$\pm$0.07 & 24 & 16.70 & He-sdOB \\
335.344036$^{*}$ & 5.416204 & 75501161$^{\ddagger}$ & 2708400230759920768 & 35340$\pm$290 & 5.86$\pm$0.07 & -0.76$\pm$0.04 & 25 & 15.28 & He-sdOB \\
335.66121$^{*}$ & 0.856942 & 382504121 & 2679230771390593024 & 50190$\pm$1050 & 5.55$\pm$0.08 & -3.09$\pm$0.15 & 29 & 16.31 & sdO \\
337.765186$^{*}$ & 12.285902 & 157014157 & 2730881326537936896 & 41950$\pm$850 & 5.40$\pm$0.05 & -3.00> & 25 & 15.69 & sdO \\
338.731397 & 31.39054 & 1203114 & 1900710357773768448 & 47320$\pm$520 & 5.93$\pm$0.50 & 0.47$\pm$0.06 & 19 & 16.47 & He-sdOB \\
339.806068$^{*}$ & 13.636279 & 157011113 & 2731836527264665856 & 32200$\pm$400 & 5.66$\pm$0.03 & -2.88> & 65 & 14.93 & sdB \\
339.981606$^{*}$ & -0.877811 & 362205243 & 2653065864983035392 & 42060$\pm$230 & 5.45$\pm$0.03 & 1.20$\pm$0.06 & 30 & 15.07 & He-sdOB \\
340.0594$^{*}$ & 2.10893 & 362211023 & 2654860474117747200 & 28280$\pm$210 & 5.38$\pm$0.05 & -2.58$\pm$0.11 & 19 & 15.90 & sdB \\
340.0884 & 12.09255 & 379902114 & 2729936846049609728 & 30670$\pm$290 & 5.42$\pm$0.07 & -3.00> & 25 & 16.76 & sdB \\
341.7299$^{*}$ & 15.5164 & 354509102 & 2732552133240294272 & 41210$\pm$40 & 5.51$\pm$0.04 & 1.75$\pm$0.07 & 29 & 15.72 & He-sdOB \\
343.076508$^{*}$ & 22.301397 & 353313201 & 2836713031563656704 & 47740$\pm$630 & 5.35$\pm$0.09 & 1.31$\pm$0.29 & 26 & 16.03 & He-sdOB \\
343.67627 & 10.23328 & 380403222 & 2717973090092215168 & 34240$\pm$810 & 5.38$\pm$0.03 & -2.65$\pm$0.14 & 26 & 16.79 & sdB \\
346.361835 & 19.298441 & 381310006 & 2831429843832458752 & 28060$\pm$160 & 5.17$\pm$0.06 & -2.60$\pm$0.17 & 10 & 17.04 & sdB \\
347.558118$^{*}$ & 34.032284 & 264102247 & 1911572540520290048 & 49490$\pm$830 & 5.62$\pm$0.08 & 0.50$\pm$0.15 & 49 & 15.44 & He-sdOB \\
348.4784 & 7.21383 & 387903096 & 2665196154002229376 & 43610$\pm$620 & 5.65$\pm$0.15 & -2.64$\pm$0.24 & 10 & 16.95 & sdO \\
350.293029 & 19.24195 & 381306077 & 2825005328672819072 & 28730$\pm$240 & 5.46$\pm$0.04 & -2.56$\pm$0.10 & 22 & 16.69 & sdB \\
351.527451$^{*}$ & 5.270987 & 158603071 & 2661097763064489600 & 29340$\pm$70 & 5.63$\pm$0.02 & -2.76$\pm$0.06 & 64 & 15.41 & sdB \\
359.439198$^{*}$ & 24.43959 & 492005208 & 2851511427281425536 & 53970$\pm$6490 & 5.61$\pm$0.07 & -2.40$\pm$0.19 & 12 & 16.36 & sdO \\
\hline\noalign{\smallskip} 
  \end{tabularx}
\end{table*}

By using the method described in Section 3, we identified 388 
hot subdwarfs in this study. Based on the classification scheme of Paper I,  
186 sdB, 73 He-sdOB, 65 sdOB,  45 sdO, 12 He-sdO and 7 He-sdB stars 
are classified, respectively. We also cross-matched our hot subdwarfs 
with the hot subdwarfs cataloged by Geier et al. (2017a), N\'emeth et al. (2012), 
and Luo et al. (2016), and got 253, 12 and 50 common stars, respectively. Note that 
nearly all the hot subdwarfs in  N\'emeth et al. (2012) and Luo et al. (2016) are 
cataloged by Geier et al. (2017a). It means that 135 new hot subdwarf stars are found 
in this study, which have not been cataloged before. Furthermore, Among the 253 common stars 
with Geier et al. (2017a), only 91 stars have their parameters available in the catalog, which are 
mostly taken from N\'emeth et al. (2012) and Luo et al. (2016). 

The parameters of the 388 hot subdwarf stars are listed in Table 1. 
Columns 1-4 give the right ascension (RA), declination (DEC), LAMOST\_obsid and Gaia source\_id. Next, columns 5-7 give the $T_{\rm eff}$, $\log{g}$ and $\log(n{\rm He}/n{\rm H})$ fitted by {\sc XTgrid}. 
Columns 8-10 list the SNR in the $u$ band, the apparent magnitude in 
the Gaia $G$ band and the spectral classification, respectively. 
The common stars  in  Geier et al. (2017a)  are marked with $^{*}$ in 
Table 1,  and the common stars with N\'emeth et al. (2012) are 
marked by $^{\dagger}$, while the common stars with Luo et al. (2016) are 
marked by $^{\ddagger}$. In table 1, 
the symbol  '$>$' in $\mathrm{log}(n\mathrm{He}/n\mathrm{H})$ 
denotes an upper limit of the He abundance, when 
{\sc XTgrid} could not find the error bars mostly due to the low quality of the spectra.  

\subsection{Parameter diagrams}
\begin{figure}
\centering

\begin{minipage}[c]{0.42\textwidth}
\includegraphics [width=75mm]{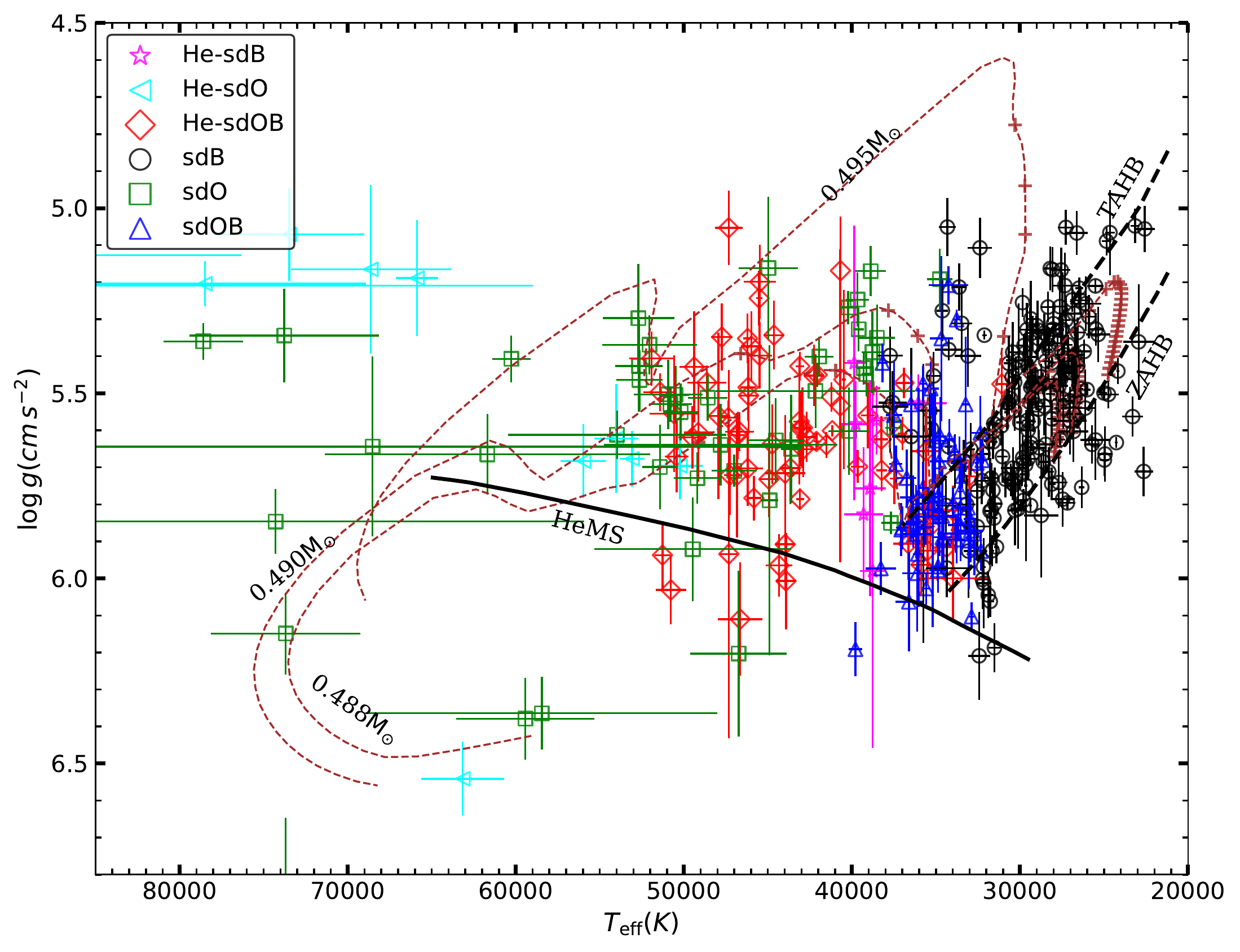}
\centerline{(a) }
\end{minipage}%
\begin{minipage}[c]{0.42\textwidth}
\includegraphics [width=75mm]{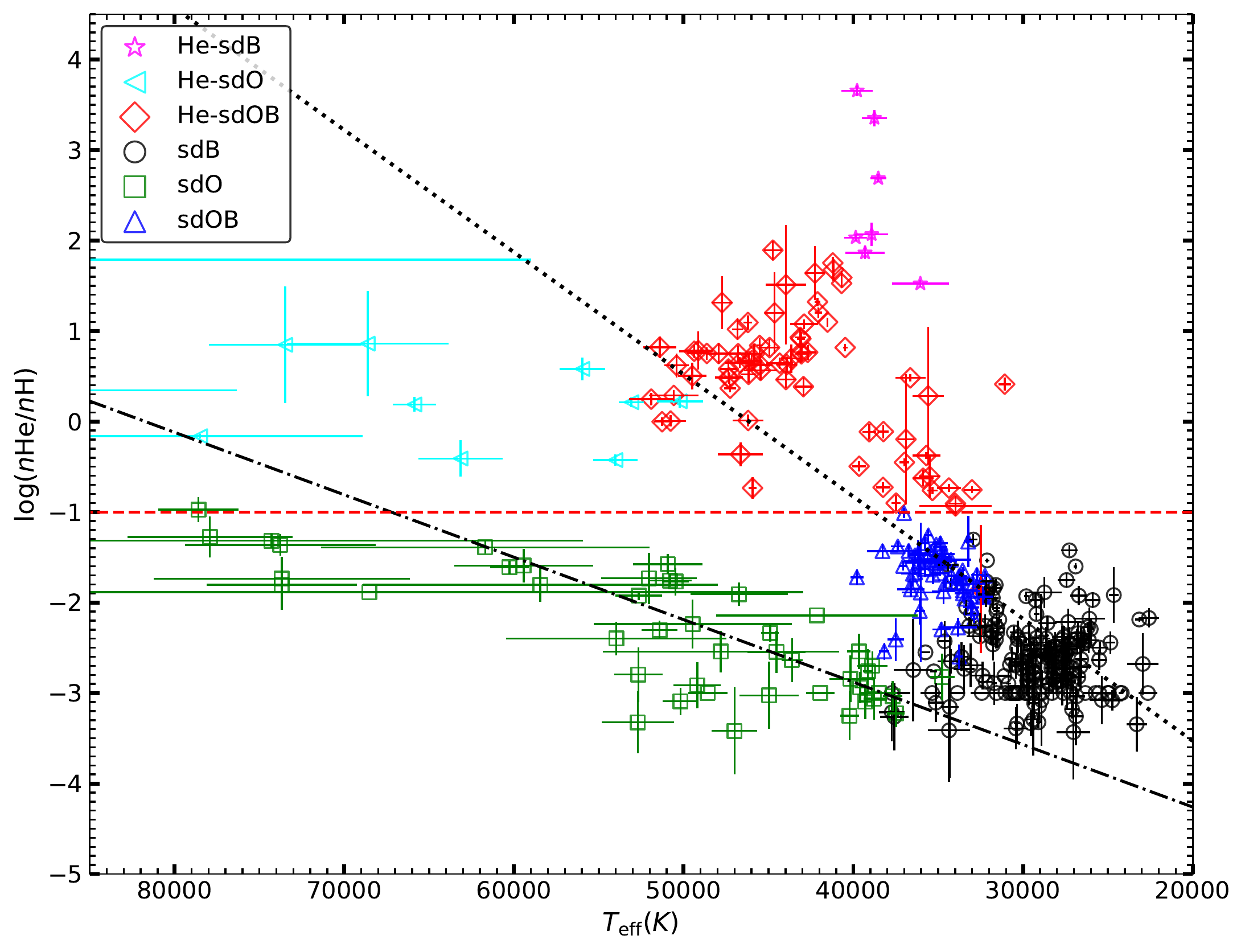} 
\centerline{(b) }
\end{minipage}\\%
\begin{minipage}[c]{0.45\textwidth}
\includegraphics [width=75mm]{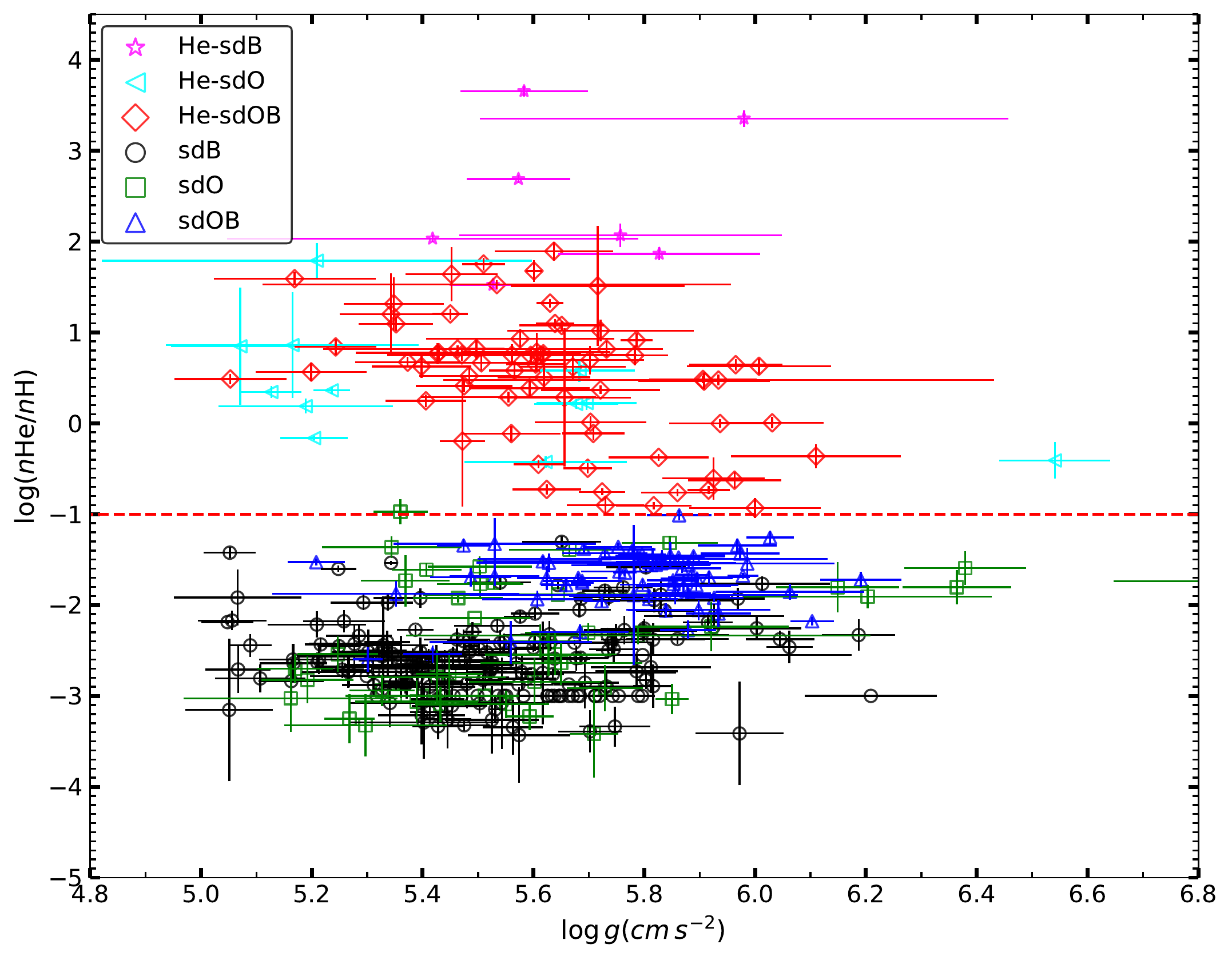} 
\centerline{(c) }
\end{minipage}%

\caption{Panel (a): $T_{\rm eff}$-$\log{g}$ diagram for the 388  
hot subdwarf stars identified in this study. The ZAHB and TAHB sequences 
with [Fe/H]= -1.48 from Dorman et al. (1993) are denoted by dashed lines. 
The He-MS from Paczy\'nski (1971) is marked by black solid line. 
Three evolution tracks for hot HB stars from Dorman et al. (1993) 
are showed with brown dotted curves (see the text for details). 
Panel (b): $T_{\rm eff}$-$\mathrm{log}(n\mathrm{He}/n\mathrm{H})$ 
diagram for the 388 hot subdwarf stars identified in this study. Different 
spectral classification types of stars are represented  by different markers 
(see the text for details). 
The black dotted line and dot-dashed line are the linear regression lines fitted by 
 by Edelmann et al. (2003) and N\'emeth et al. (2012), respectively, while 
 the red dashed line marks the solar value of He abundance. Panel (c): 
$\log{g}$-$\mathrm{log}(n\mathrm{He}/n\mathrm{H})$ diagram for the 
388 hot subdwarf stars identified in this study.} 
\end{figure}

Panel (a), (b) and (c) in Fig 4 show the $T_{\rm eff}$-$\log{g}$, $T_{\rm eff}$-$\mathrm{log}(n\mathrm{He}/n\mathrm{H})$ and $\log{g}$-$\mathrm{log}(n\mathrm{He}/n\mathrm{H})$ diagram 
for the 388 identified hot subdwarfs, respectively. 
The He-sdB, He-sdO, He-sdOB, sdB,  sdO and sdOB stars are 
marked by magenta stars, aqua left triangles, red diamonds, 
black circles, green squares and blue up triangles, respectively. 
The zero-age HB (ZAHB)  and terminal-age (TAHB)  with [Fe/H]=-1.48  
from Dorman et al. (1993) are represented by two dashed lines in Panel (a), 
while the He-MS from Paczy\'nski (1971) is denoted by a black solid line. 
We also show three  evolution tracks for hot HB stars by three brown dotted curves with 
masses from top to bottom: 0.495, 0.490 and 0.488, respectively. 
The  dotted line in Panel (b) is the linear regression line used to 
fit the He-rich sequence  by Edelmann et al. (2003), 
while the dot-dashed line is the regression line used to fit 
the He-weak sequence by N\'emeth et al. (2012). The red 
horizontal dashed line in Panel (b) and (c) denotes the solar He abundance (log($n{\rm He}/n{\rm H}) = -1$). 

Panel (a) of Fig 4 shows that most of the sdB (i.e., black circles) and 
sdOB stars (i.e., blue up triangles) are well  in the region defined by the ZAHB and TAHB lines, 
which indicates that these stars are He core burning stars. 
In addition,  our sdOB stars present higher temperatures and gravity than sdB stars. 
On the other hand, most He-sdOB stars (red diamonds) evolve off the TAHB 
and cluster near  $T_{\rm eff}$=45\,000 K and 
$\log{g}$=5.6 $\mathrm{cm\ s^{-2}}$, but we also find a few He-sdOB stars  
in the bottom areas defined by the ZAHB and TAHB, and overlap with our sdOB stars. 
Both  sdO stars (i.e., green squares) and He-sdO stars (i.e., aqua left triangles) 
show very high temperatures (e.g., $T_{\rm eff}>$40\,000 K), and some of them even 
have their temperatures over 70\,000 K. Moreover, the gravity of these two groups  
also cover a wide range. 7 He-sdB stars are found in our sample, which are marked 
by magenta stars in Fig 4. These stars have the temperature around 40\,000 K, but a 
wide coverage of gravity. 

Two distinct He sequences (e.g., He-rich sequence fitted by the 
dotted line (Edelmann et al. 2003) and He-weak sequence fitted by 
the dot-dashed line (N\'emeth et al. 2012)) are clearly presented in Panel (b) . 
This obvious characteristics  
in field hot subdwarfs was first found by (Edelmann et al. 2003) and 
confirmed later by many other authors (N\'emeth et al. 2012; Geier et al. 2013; 
Luo et al. 2016; Paper I). One can see from Panel (b), the He-rich sequence consists of 
sdB, sdOB, He-sdOB and He-sdB stars, while the He-weak sequence consists of 
all the sdO stars and several sdB stars with very low He abundances  
(e.g., $\mathrm{log}(n\mathrm{He}/n\mathrm{H})<$-3.0). However, 
due to the low resolution of the LAMOST spectra  (e.g., $\lambda/\Delta\lambda$=1800), 
the He abundances for some of  He-poor sdB stars are difficult to obtain or 
are obtained with very large uncertainties (e.g., $\mathrm{log}(n\mathrm{He}/n\mathrm{H})<$-3.5). 
For such objects we report  an upper limit of $\mathrm{log}(n\mathrm{He}/n\mathrm{H})=$-3.0 in this study. 
Considering this influence, one would expect that there could be more sdB stars 
make up the He-weak sequence in Panel (b). 
On the other hand, most He-sdO stars in our sample (i.e., aqua left triangles)  
are located in the region between the two regression lines, but much closer to the He-weak sequence. 
More interestingly, the He-sdOB stars in our sample (i.e., red diamonds) are split into two 
sub-groups by a distinct gap  at $T_{\rm eff}$=40\,000 K and 
$\mathrm{log}(n\mathrm{He}/n\mathrm{H})$=0.0. This distinct gap 
is also present in other studies (e.g., Fig 6 in N\'emeth et al. 2012;  
Fig 8 in Luo et al. 2016;  Fig 4 and Fig 5 in Heber 2016; Fig 7 in Paper I). 
Further work using higher resolution data and more complex models are needed for this feature, 
a detailed study on this aspect is out the scope of this paper.  
The distribution of hot subdwarf stars in our sample covers  a wide range of gravity in Panel (c). 

\subsection{Comparison with other studies} 
\begin{figure}
\centering

\begin{minipage}[c]{0.70\textwidth}
\includegraphics [width=120mm]{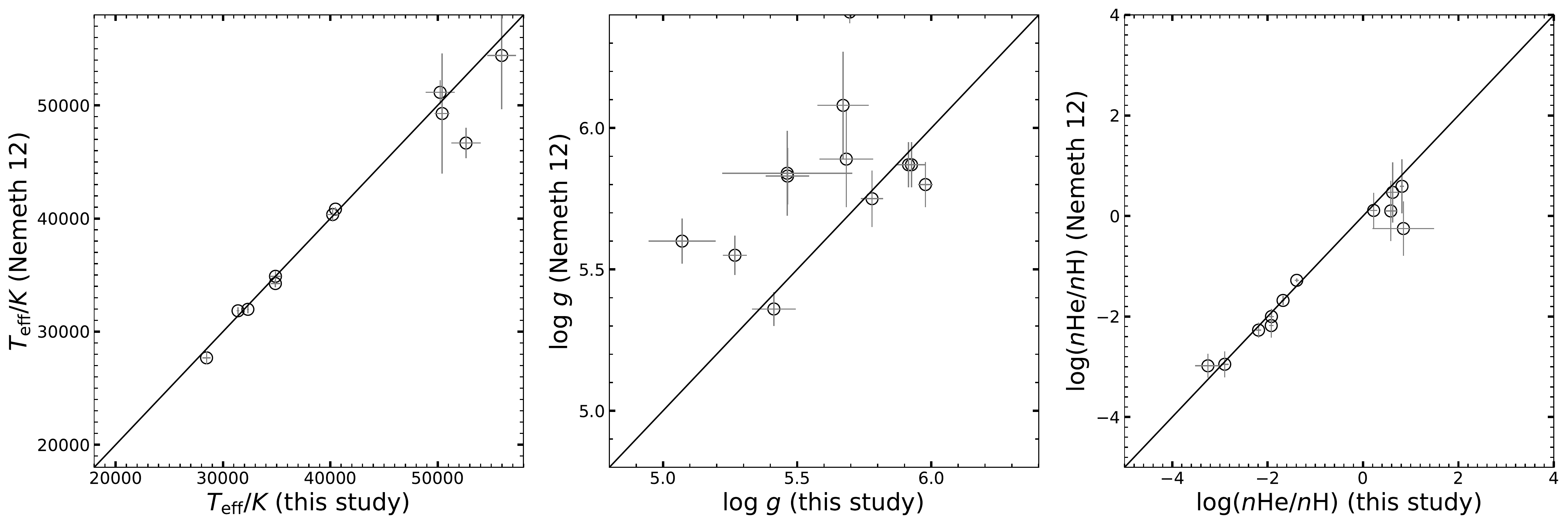} 
\centerline{(a) }
\end{minipage}\\%
\begin{minipage}[c]{0.70\textwidth}
\includegraphics [width=120mm]{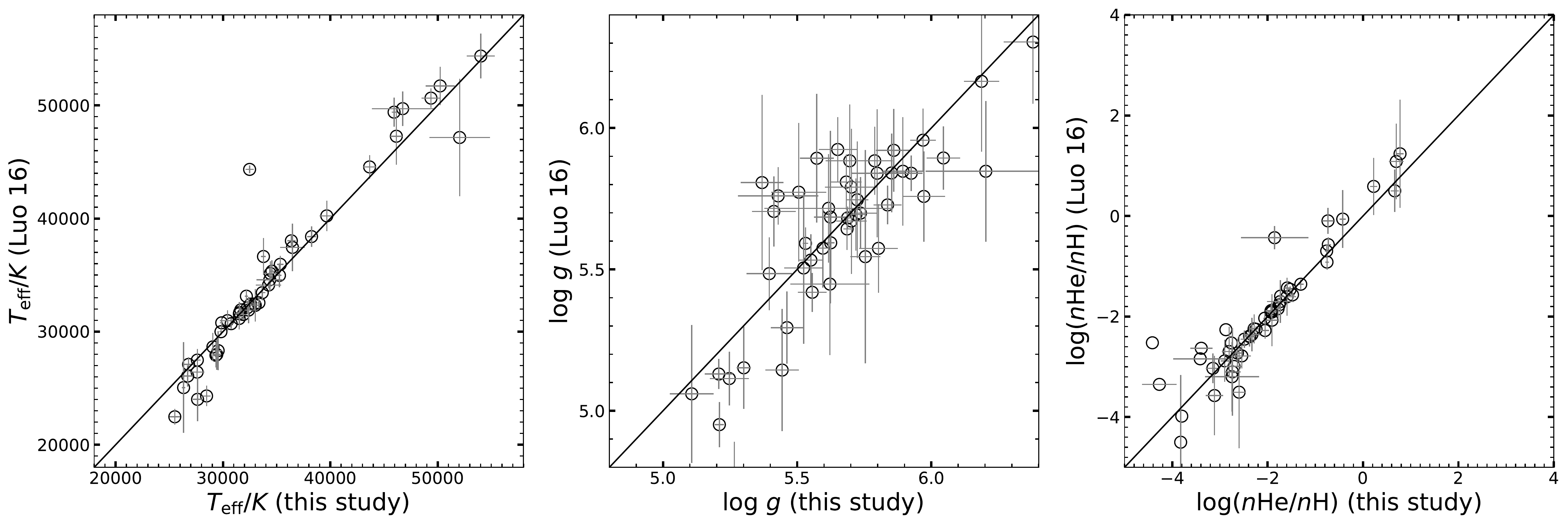} 
\centerline{(b) }
\end{minipage}\\%

\caption{
Panel (a): Parameter comparison with  N\'emeth et al. (2012). 
Panel (b): Parameter comparison with Luo et al. (2016).}
\end{figure}

Like in Paper I, we also compared  the atmospheric parameters of the stars that are common in our 
study and other studies, where the atmospheric parameters are available. 
Panel (a) and (b) in Fig 5 show the results from a comparison of our study 
and in N\'emeth et al. (2012),   Luo et al. (2016), respectively. 
One can see that the $T_{\rm eff}$ and $\mathrm{log}(n\mathrm{He}/n\mathrm{H})$ 
in our study matched well with the values from the two other studies. However, a larger dispersion is 
present in gravity than in the other two parameters in Fig 5.  Note that, the hot subdwarf  spectra in  
N\'emeth et al. (2012) are obtained in the {\it GALEX survey}, which have different quality (e.g., SNR) 
from our LAMOST spectra. Further more, the atmospheric models used in N\'emeth et al. (2012) 
contain H, He, C, N and O compositions, while the atmospheric models used in this study 
contain H and He compositions only. On the other hand, although the atmospheric models and 
fitting procedure used in Luo et al. (2016) are the same with the ones we used in this study, but 
the spectra used for spectral analysis also could be different. Luo et al. (2016) selected 
candidates from LAMOST DR1, while we selected candidates from LAMOST DR5. The 
spectra could be different in SNR even for the same objects due to the repeat observations in LAMOST. These 
facts can result in the large gravity discrepancy between these two studies and our results. 
Nevertheless, the values of $\log{g}$ in our study are comparable to the ones in other studies, and 
the comparison results showed in Fig 5 demonstrate the reliability of the spectral analysis in this study.  

\section{Discussion}  
The counterparts of field hot subdwarf stars  in GCs  
are the EHB  and blue hook (BHk) stars (Heber 2009, 2016; D'cruz et al. 2000; 
Brown et al. 2001, 2016, Latour et al. 2014). 
Although these stars are burning He in the cores or even have  
evolved off the TAHB, the origin of the two counterparts could be 
different due to the very different environments, ages and  initial compositions  
(for some formation scenarios of the two counterparts see Han et al. 2002, 2003; 
Chen et al. 2013; Zhang et al. 2012, 2017; Lei et al. 2015, 2016; Xiong et al. 2017; Wu et al. 2018). 

Latour et al. (2018) analyzed the spectra of 152 EHB stars in $\omega$ Cen, 
which were obtained with the FORS1 and VIMOS 
spectrographs equipped on the Very Large Telescope (VLT, ESO).  
The atmospheric parameters (e.g., $T_{\rm eff}$, $\log{g}$ and  
$\mathrm{log}(n\mathrm{He}/n\mathrm{H})$) are derived by 
adopting non-LTE model atmospheres. Latour et al. (2018) found  three 
distinct groups presented in their sample, e.g., H-sdB stars (the coolest 
H-rich stars), H-sdO stars (the hottest H-rich stars) and He-sdOB stars
(see Fig 5 in their study). Moreover, the He-sdOB group can be 
further sub-divided into two sub-groups based on their He abundances. 
Since the sample size of the analyzed EHB stars in $\omega$ Cen is big 
enough, Latour et al. (2018) also compared their results with 
four representative samples of field hot subdwarfs (i.e., Edelmann et al. 2003; 
Lisker et al. 2005; Stroeer et al. 2007; N\'emeth et al. 2012). 
They found that the two distinct He sequences present in 
field hot subdwarfs also appear in $\omega$ Cen EHB stars. 
However, the number fractions among different groups  
of EHB stars in $\omega$ Cen are very different from 
the ones found in the field. 
The He-sdOB populations (corresponding to BHk stars) in $\omega$ Cen are less 
represented in field hot subdwarf stars (e.g., $\approx$ 5\% 
in the galactic disk, and   $\approx$ 23\% in the galactic halo, while 
 $\approx$ 52\% in $\omega$ Cen; Geier et al. 2017b). 
 The results in Latour et al. (2018) demonstrated that BHk stars in 
 GCs could experience a different formation origin from field hot subdwarf stars. 
 
Similar results have been obtained by Heber (2016, see Fig 23 
 in their study) by comparing EHB stars in $\omega$ Cen and 
 NGC 2808 with the field hot subdwarf stars from the ESO-SPY project. They also concluded that 
 He-rich hot subdwarf stars in the field are hotter than the He-rich 
 EHB stars in GCs, and the He enrichment is higher in many field 
 stars than that of the most He enriched stars in GCs. 
 
\begin{figure}
\centering
\includegraphics [width=140mm]{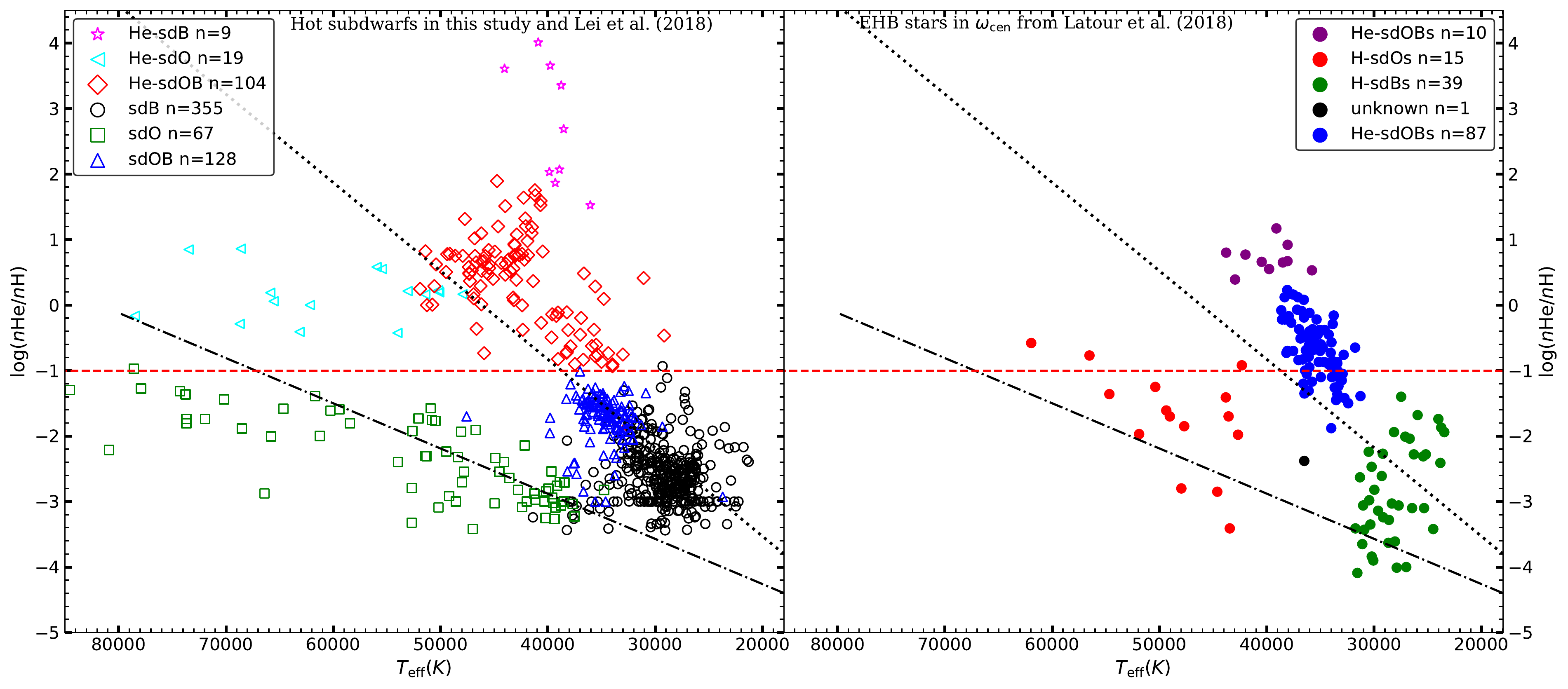}
\caption{Comparing the 682 hot subdwarf stars identified in this study and Paper I 
with the $\omega$ Cen EHB stars  analyzed by Latour et al. (2018) in the 
$T_{\rm eff}$-$\mathrm{log}(n\mathrm{He}/n\mathrm{H})$) 
diagram. The different markers and lines in the left panel have 
the same meanings as that of in Panel (b) of Fig 4. The solid circles with different 
colors in the right panel denote the different EHB groups 
found in Latour et al. (2018, see text for details). The error bars 
are not showed for clarity.}
\end{figure}
 
To have a more clear understanding on this problem, we also compared 
the field  hot subdwarf stars identified in this study and Paper I (682 stars in total, 
including 355 sdB, 128 sdOB, 104 He-sdOB, 67 sdO, 19 He-sdO and 9 He-sdB stars) with 
the EHB stars in $\omega$ Cen analyzed by Latour et al. (2018). 
Fig 6 shows the comparison in the $T_{\rm eff}$-$\mathrm{log}(n\mathrm{He}/n\mathrm{H})$) 
diagram between our field hot subdwarfs and the EHB stars in $\omega$ Cen. 
The red dashed line marks the solar He abundance (i.e., $\mathrm{log}(n\mathrm{He}/n\mathrm{H})$ = -1). 
The dotted and dot-dashed lines are the same regression lines marked in Panel (b) of Fig 4, which are used to 
fit the He-rich sequence and He-weak sequence by Edelmann et al. (2003) and N\'emeth et al. (2012), 
respectively. The 682 hot subdwarfs in our sample are denoted in the left panel  by the same markers as in Fig 4, 
while the 152 EHB stars in $\omega$ Cen are denoted in the right panel by solid circles with different colors. 

As described by Latour et al. (2018), there are three distinct groups of  EHB stars in $\omega$ Cen 
\footnote{One star in Latour et al. (2018) which can be added into any groups is marked with a black solid circle.}. 
The H-sdB stars in $\omega$ Cen (i.e., green solid circles in the right panel) 
correspond to our sdB stars (i.e., black circles in the left panel). These stars 
have the lowest He abundance and temperatures (e.g., $\mathrm{log}(n\mathrm{He}/n\mathrm{H})<$ -2 and 
$T_{\rm eff}<$ 32\,000 K) among the three groups. The He-sdOB group in $\omega$ Cen can be sub-divided into 
two sub-groups, e.g., a high He abundance group (i.e., purple solid circles in the right panel) 
and a low He abundance group (i.e., blue solid circles). This group corresponds to our He-sdOB stars 
(i.e., red diamonds in the left panel). As we described in Section 4, our He-sdOB stars can be divided 
into two distinct groups based on their He abundances as well. Furthermore, 
the He-sdOB sub-group with low He  in $\omega$ Cen is much bigger 
in size than the He-sdOB sub-group with higher He (e.g., 87 vs 10, or the number fraction of 90\% vs 10\%). 
However, it is the opposite  case in our He-sdOB samples, that the low He 
group is much smaller in size than the high He group (e.g. 33 vs 71, 
or the number fraction of 32\% vs 68\%) in our sample.  Moreover, similar to the conclusion in  Heber (2016), 
our He-sdOB stars in the high He group present higher temperatures 
(e.g., most of them have $T_{\rm eff}>$ 40\,000 K) than the high He He-sdOB stars 
in $\omega$ Cen (e.g, many of them have  $T_{\rm eff}<$ 40\,000 K).  
There is a distinct gap between H-sdB stars and He-sdOB stars 
in $\omega$ Cen (see the right panel). As mentioned by Latour et al. (2018) 
this gap is predicted by the normal EHB models and delayed  He-flash models 
(Miller Bertolami et al. 2008; Moehler et al. 2011, Lei et al. 2015, 2016). 
However, this distinct gap is not present in our sample (see the left panel), and 
it seems to be filled up by some sdB stars and sdOB stars in the same region. 
The H-sdO stars in $\omega$ Cen (i.e., red solid circles in the right panel) correspond to 
our sdO stars and He-sdO stars (e.g., green open squares and aqua left triangles in the 
left panel). Although the He-rich stars belong to this group  of $\omega$ Cen 
(e.g., $\mathrm{log}(n\mathrm{He}/n\mathrm{H})>$ -1) 
are much less in numbers than that of in our sample (e.g., 3 vs 19), but 
the number fraction are comparable for the two groups 
(e.g., 21\% in our sdO and He-sdO samples vs 20\% in $\omega$ Cen H-sdO stars). In the left panel, 
we find 9 He-sdB stars (i.e., magenta stars), which have 
the highest He abundance (e.g., $\mathrm{log}(n\mathrm{He}/n\mathrm{H})>$ 1) 
in our sample, but this group is missing in $\omega$ Cen. 

 \begin{figure}
\centering
\includegraphics [width=140mm]{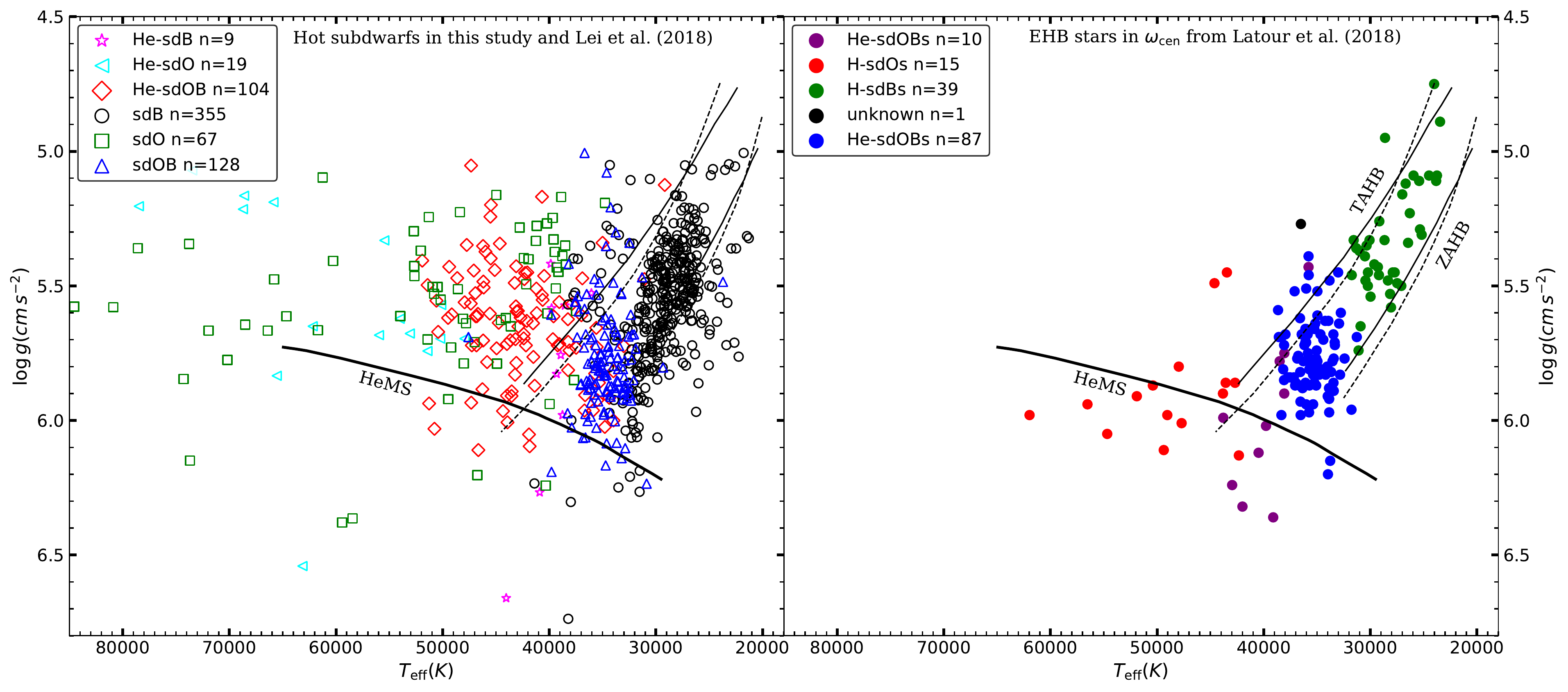}
\caption{Comparing the 682 hot subdwarf stars identified in this study and Paper I 
with the $\omega$ Cen EHB stars  analyzed by Latour et al. (2018) in the 
$T_{\rm eff}$-$\log{g}$ diagram. The different markers and lines in the left panel  
have the same meanings as that of in Panel (a) of Fig 4. The solid circles with 
different colors in the right panel denote the different EHB groups found in Latour et al. (2018, 
see text for details). The error bars are not showed for clarity.  }
\end{figure}

Fig 7 shows the comparison of  our hot subdwarfs and the 
$\omega$ Cen EHB stars  in $T_{\rm eff}$-$\log{g}$ 
diagram. The two dashed lines are the ZAHB and TAHB sequences with 
$Y$ = 0.24 from Latour et al. (2018), while  the two thin solid lines are 
the ZAHB and TAHB sequences with $Y$ = 0.40 from Latour et al. (2018). The thick solid line is  
the He-MS from Paczy\'nski (1971). One can see that the gap between the H-sdB stars 
and He-sdOB stars in $\omega$ Cen discussed above shows up clearly in the right panel of Fig 7. 
This gap is filled up by sdB stars and a few sdOB stars in our 
sample (see the left panel).  
The comparison between our hot subdwarfs and the $\omega$ Cen EHB stars in the 
$\log{g}$-$\mathrm{log}(n\mathrm{He}/n\mathrm{H})$ diagram is showed 
in Fig 8. The panel on the right reveals a a strong positive correlation between 
gravity and He abundance  in $\omega$ Cen. However, this correlation 
appears more obscure in the field (see the left panel). 

\begin{figure}
\centering
\includegraphics [width=140mm]{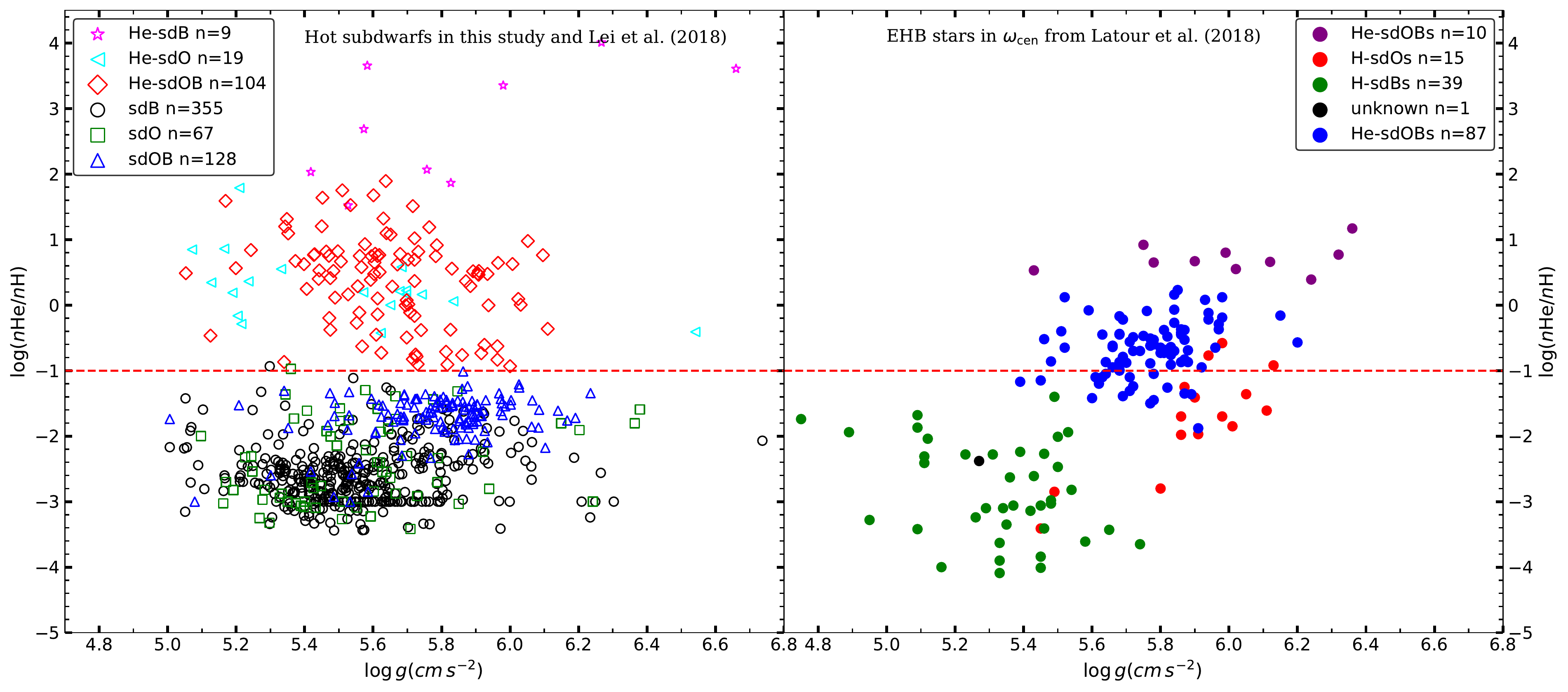}
\caption{Comparing the 682 hot subdwarf stars identified in this study and Paper I  
with the $\omega$ Cen EHB stars  analyzed by Latour et al. (2018) in the 
$\log{g}$-$\mathrm{log}(n\mathrm{He}/n\mathrm{H})$ diagram. 
The different markers and lines in the left panel  
have the same meanings as that of in Panel (c) of Fig 4. The solid circles with 
different colors in the right panel denote the different EHB groups found by Latour et al. (2018, 
see text for details). The error bars are not showed for clarity. } 
\end{figure}

The comparison results presented in this section come to a similar conclusion as in  
Latour et al. (2018) and Heber (2016) that the formation origins of field 
hot subdwarf stars and EHB stars in GCs are likely different in some aspects. 
Detailed theoretical  models are needed to  uncover the underlying differences of these populations. 
The two sub-groups of He-sdOB stars in the $T_{\rm eff}$-$\mathrm{log}(n\mathrm{He}/n\mathrm{H})$ 
diagram which are clearly present in the 
EHB stars of $\omega$ Cen also appear in our LAMOST field sample. However, 
this feature is not present in the ESO-SPY sample of field subdwarfs  (see lower left hand panel of Fig 23 
in Heber et al. 2016). This could be due to the fact that the size of the He-sdOB group in 
ESO-SPY sample is much smaller than that of our LAMOST sample.

\section{Summary} 
We have selected 2074 hot subdwarf candidates in the HR-diagram which was 
built by cross-matching the Gaia DR2 database with the LAMOST DR5 spectral database. 
After conducting a detailed spectral analysis, we identified 388 hot subdwarf stars 
among 441 candidates. The atmospheric parameters have been derived from non-LTE  
model atmospheres, and 186 sdB, 73 He-sdOB, 65 sdOB,  45 sdO, 12 He-sdO and 7 He-sdB stars were found in 
our study. Together with the 294 hot subdwarf stars found in Paper I, we totally 
identified 682 hot subdwarf  stars by combining the Gaia  database with the 
LAMOST database, among which, 241 hot subdwarf stars are  
newly identified that were not cataloged before. While 441 stars have records 
in the catalog of Geier et al. (2017a), but 255 of them have no 
atmospheric parameters. It means that we not only 
identified 241 new hot subdwarf stars, 
but also newly obtained atmospheric parameters for 255 hot subdwarf stars 
which were already cataloged in Geier et al. (2017a).
These results indicate the efficiency of our  method 
to select hot subdwarf candidates by cross-matching these two large survey databases. We 
expect that a large number of new hot subdwarfs will be discovered with the new data release of 
these surveys, which will make great contributions on the study of these special blue stars. 

We confirmed the two distinct He sequences in the 
$T_{\rm eff}$-$\mathrm{log}(n\mathrm{He}/n\mathrm{H})$ diagram. In addition, 
we found an obvious gap in our He-sdOB stars, which is also present in 
$\omega$ Cen EHB stars. However, the number fraction of the members in the two sub-groups 
is very different  between field samples and the $\omega$ Cen EHB stars.  
Furthermore, the distinct gap between the H-sdB group and the He-sdOB group 
in $\omega$ Cen EHB stars is not present in our sample. More interestingly, 
the He-sdB group which have the highest He abundance in our sample is 
completely missing in the $\omega$ Cen. These results indicate a different origin of  
the field hot subdwarf stars and their GC counterparts.

\acknowledgments
We thank the anonymous referee for his/her valuable suggestions and 
comments which improve this work greatly.  
This work is supported by the National Natural Science Foundation 
of China Grant Nos 11503016 and 11390371,   
Natural Science Foundation of Hunan province Grant No.2017JJ3283,  
the Youth Fund project of Hunan Provincial Education Department
Grant No.15B214, the Astronomical Big Data Joint Research
Center, co-founded by the National Astronomical Observatories, Chinese 
Academy of Sciences and the Alibaba Cloud. 
This research has used the services of \mbox{\url{www.Astroserver.org}} under reference XD8O2C. 
P.N. acknowledges support from the Grant Agency of the Czech Republic (GA\v{C}R 18-20083S). 
This project was developed in part at the 2018 
Gaia-LAMOST Sprint workshop, supported by the NSFC under
grants 11333003 and 11390372. 
The LAMOST Fellowship is supported by Special Funding for Advanced Users, 
budgeted and administered by the Center for Astronomical 
Mega-Science, Chinese Academy of Sciences (CAMS). 
Guoshoujing Telescope (the Large Sky Area Multi-Object Fiber 
Spectroscopic Telescope LAMOST) is a National Major Scientific 
Project built by the Chinese Academy of Sciences. 
Funding for the project has been provided by the 
National Development and Reform Commission. 
LAMOST is operated and managed by the National Astronomical Observatories, 
Chinese Academy of Sciences.



\end{document}